\pdfoutput=1
\documentclass[12pt,a4paper]{article}

\usepackage{ifthen} 
\newboolean{pdflatex}
\setboolean{pdflatex}{true} 

\newboolean{articletitles}
\setboolean{articletitles}{true} 

\newboolean{uprightparticles}
\setboolean{uprightparticles}{false} 

\def\paperauthors{LHCb collaboration} 
\def\paperasciititle{First measurement of the CP-violating phase in Bs->J/psi(e+e-)phi decays} 
\def\papertitle{First measurement\\ of the \CP-violating phase\\ in \decay{\Bs}{\jpsi(}{\to\epem)\phiz} decays} 
\def\paperkeywords{{High Energy Physics}, {LHCb}} 
\def\papercopyright{\the\year\ CERN for the benefit of the LHCb collaboration} 
\def\paperlicence{CC BY 4.0 licence}
\def\paperlicenceurl{https://creativecommons.org/licenses/by/4.0/}

\usepackage[top=1in, bottom=1.25in, left=1in, right=1in]{geometry}

\usepackage{microtype}
\usepackage{lineno}  
\usepackage{xspace} 
\usepackage{caption} 

\usepackage{graphicx}  
\usepackage{color}
\usepackage{colortbl}
\graphicspath{{./figs/}} 

\usepackage{amsmath} 
\usepackage{amssymb}
\usepackage{amsfonts}
\usepackage{upgreek} 

\newcommand*\patchAmsMathEnvironmentForLineno[1]{%
\expandafter\let\csname old#1\expandafter\endcsname\csname #1\endcsname
\expandafter\let\csname oldend#1\expandafter\endcsname\csname
end#1\endcsname
 \renewenvironment{#1}%
   {\linenomath\csname old#1\endcsname}%
   {\csname oldend#1\endcsname\endlinenomath}%
}
\newcommand*\patchBothAmsMathEnvironmentsForLineno[1]{%
  \patchAmsMathEnvironmentForLineno{#1}%
  \patchAmsMathEnvironmentForLineno{#1*}%
}
\AtBeginDocument{%
\patchBothAmsMathEnvironmentsForLineno{equation}%
\patchBothAmsMathEnvironmentsForLineno{align}%
\patchBothAmsMathEnvironmentsForLineno{flalign}%
\patchBothAmsMathEnvironmentsForLineno{alignat}%
\patchBothAmsMathEnvironmentsForLineno{gather}%
\patchBothAmsMathEnvironmentsForLineno{multline}%
\patchBothAmsMathEnvironmentsForLineno{eqnarray}%
}

\usepackage{hyperxmp}

\usepackage[pdftex,
            pdfauthor={\paperauthors},
            pdftitle={\paperasciititle},
            pdfkeywords={\paperkeywords},
            pdfcopyright={Copyright (C) \papercopyright},
            pdflicenseurl={\paperlicenceurl}]{hyperref}

\usepackage[colorinlistoftodos,textsize=scriptsize]{todonotes}

\usepackage[bottom,flushmargin,hang,multiple]{footmisc}

\usepackage[all]{hypcap} 

\usepackage{xspace} 
\usepackage{upgreek}

\def\lhcb   {\mbox{LHCb}\xspace}
\def\atlas  {\mbox{ATLAS}\xspace}
\def\cms    {\mbox{CMS}\xspace}

\def\cdf    {\mbox{CDF}\xspace}
\def\dzero  {\mbox{D0}\xspace}

\def\lhc    {\mbox{LHC}\xspace}

\def\MagUp {\mbox{\em Mag\kern -0.05em Up}\xspace}

\ifthenelse{\boolean{uprightparticles}}%
{
 
 \def\Pgamma      {\ensuremath{\upgamma}\xspace}

 \def\Pmu         {\ensuremath{\upmu}\xspace}

 \def\Ppi         {\ensuremath{\uppi}\xspace}

 \def\Pphi        {\ensuremath{\upphi}\xspace}                 
                  
 \def\Pchi        {\ensuremath{\upchi}\xspace}                 
 \def\Ppsi        {\ensuremath{\uppsi}\xspace}

 \def\PDelta      {\ensuremath{\Delta}\xspace}                 
 \def\PXi         {\ensuremath{\Xi}\xspace}                 
 \def\PLambda     {\ensuremath{\Lambda}\xspace}                 
 \def\PSigma      {\ensuremath{\Sigma}\xspace}                 
 \def\POmega      {\ensuremath{\Omega}\xspace}                 
 \def\PUpsilon    {\ensuremath{\Upsilon}\xspace}

 \def\PB      {\ensuremath{\mathrm{B}}\xspace}                 
                  
 \def\PD      {\ensuremath{\mathrm{D}}\xspace}

 \def\PJ      {\ensuremath{\mathrm{J}}\xspace}                 
 \def\PK      {\ensuremath{\mathrm{K}}\xspace}

 \def\Pb      {\ensuremath{\mathrm{b}}\xspace}                 
 \def\Pc      {\ensuremath{\mathrm{c}}\xspace}                 
                  
 \def\Pe      {\ensuremath{\mathrm{e}}\xspace}

 \def\Pi      {\ensuremath{\mathrm{i}}\xspace}

 \def\Ps      {\ensuremath{\mathrm{s}}\xspace}                 
 \def\Pt      {\ensuremath{\mathrm{t}}\xspace}

 \def\thebaroffset{0.0em}
}
{
 
 \def\Pgamma      {\ensuremath{\gamma}\xspace}

 \def\Pmu         {\ensuremath{\mu}\xspace}

 \def\Ppi         {\ensuremath{\pi}\xspace}

 \def\Pphi        {\ensuremath{\phi}\xspace}                 
                  
 \def\Pchi        {\ensuremath{\chi}\xspace}                 
 \def\Ppsi        {\ensuremath{\psi}\xspace}                 
                  
 \mathchardef\PDelta="7101
 \mathchardef\PXi="7104
 \mathchardef\PLambda="7103
 \mathchardef\PSigma="7106
 \mathchardef\POmega="710A
 \mathchardef\PUpsilon="7107
                  
 \def\PB      {\ensuremath{B}\xspace}                 
                  
 \def\PD      {\ensuremath{D}\xspace}

 \def\PJ      {\ensuremath{J}\xspace}                 
 \def\PK      {\ensuremath{K}\xspace}

 \def\Pb      {\ensuremath{b}\xspace}                 
 \def\Pc      {\ensuremath{c}\xspace}                 
                  
 \def\Pe      {\ensuremath{e}\xspace}

 \def\Pi      {\ensuremath{i}\xspace}

 \def\Ps      {\ensuremath{s}\xspace}                 
 \def\Pt      {\ensuremath{t}\xspace}

 \def\thebaroffset{0.18em}
}
\newcommand{\offsetoverline}[2][\thebaroffset]{\kern #1\overline{\kern -#1 #2}}%

\makeatletter
\ifcase \@ptsize \relax
  \newcommand{\miniscule}{\@setfontsize\miniscule{4}{5}}
\or
  \newcommand{\miniscule}{\@setfontsize\miniscule{5}{6}}
\or
  \newcommand{\miniscule}{\@setfontsize\miniscule{5}{6}}
\fi
\makeatother

\DeclareRobustCommand{\optbar}[1]{\shortstack{{\miniscule (\rule[.5ex]{1.25em}{.18mm})}
  \\ [-.7ex] $#1$}}

\def\ep         {{\ensuremath{\Pe^+}}\xspace}

\def\epem       {{\ensuremath{\Pe^+\Pe^-}}\xspace}

\def\mumu       {{\ensuremath{\Pmu^+\Pmu^-}}\xspace}

\def\g      {{\ensuremath{\Pgamma}}\xspace}

\def\squark    {{\ensuremath{\Ps}}\xspace}

\def\cquark    {{\ensuremath{\Pc}}\xspace}
\def\cquarkbar {{\ensuremath{\overline \cquark}}\xspace}

\def\bquark    {{\ensuremath{\Pb}}\xspace}
\def\bquarkbar {{\ensuremath{\overline \bquark}}\xspace}

\def\tquark    {{\ensuremath{\Pt}}\xspace}

\def\pion   {{\ensuremath{\Ppi}}\xspace}
\def\piz    {{\ensuremath{\pion^0}}\xspace}
\def\pip    {{\ensuremath{\pion^+}}\xspace}
\def\pim    {{\ensuremath{\pion^-}}\xspace}

\def\kaon    {{\ensuremath{\PK}}\xspace}

\def\KorKbar {\kern \thebaroffset\optbar{\kern -\thebaroffset \PK}{}\xspace}

\def\Kp      {{\ensuremath{\kaon^+}}\xspace}
\def\Km      {{\ensuremath{\kaon^-}}\xspace}
\def\Kpm     {{\ensuremath{\kaon^\pm}}\xspace}

\def\Kstar   {{\ensuremath{\kaon^*}}\xspace}

\newcommand{\phiz}{\ensuremath{\Pphi}\xspace}

\def\D       {{\ensuremath{\PD}}\xspace}

\def\DorDbar {\kern \thebaroffset\optbar{\kern -\thebaroffset \PD}\xspace}

\def\Dp      {{\ensuremath{\D^+}}\xspace}
\def\Dm      {{\ensuremath{\D^-}}\xspace}

\def\DpDm    {\ensuremath{\Dp {\kern -0.16em \Dm}}\xspace}

\def\Dsp     {{\ensuremath{\D^+_\squark}}\xspace}
\def\Dsm     {{\ensuremath{\D^-_\squark}}\xspace}

\def\B       {{\ensuremath{\PB}}\xspace}
\def\Bbar    {{\ensuremath{\offsetoverline{\PB}}}\xspace}

\def\BorBbar {\kern \thebaroffset\optbar{\kern -\thebaroffset \PB}\xspace}

\def\Bd      {{\ensuremath{\B^0}}\xspace}

\def\BdorBdbar {\kern \thebaroffset\optbar{\kern -\thebaroffset \Bd}\xspace}

\def\Bpm     {{\ensuremath{\B^\pm}}\xspace}

\def\Bs      {{\ensuremath{\B^0_\squark}}\xspace}
\def\Bsb     {{\ensuremath{\Bbar{}^0_\squark}}\xspace}
\def\BsorBsbar {\kern \thebaroffset\optbar{\kern -\thebaroffset \Bs}\xspace}
\def\Bc      {{\ensuremath{\B_\cquark^+}}\xspace}

\def\jpsi     {{\ensuremath{{\PJ\mskip -3mu/\mskip -2mu\Ppsi}}}\xspace}
\def\psitwos  {{\ensuremath{\Ppsi{(2S)}}}\xspace}

\def\chicone  {{\ensuremath{\Pchi_{\cquark 1}}}\xspace}

\def\Y#1S{\ensuremath{\PUpsilon{(#1S)}}\xspace}

\def\Lz          {{\ensuremath{\PLambda}}\xspace}

\def\LorLbar     {\kern \thebaroffset\optbar{\kern -\thebaroffset \PLambda}\xspace}

\def\Lb           {{\ensuremath{\Lz^0_\bquark}}\xspace}

\newcommand{\decay}[2]{\ensuremath{#1\!\to #2}\xspace} 

\def\to                 {\ensuremath{\rightarrow}\xspace}

\def\eps   {{\ensuremath{\varepsilon}}\xspace}

\def\CP                {{\ensuremath{C\!P}}\xspace}

\def\Vcs  {{\ensuremath{V_{\cquark\squark}}}\xspace}
\def\Vts  {{\ensuremath{V_{\tquark\squark}}}\xspace}

\def\Vcbs  {{\ensuremath{V_{\cquark\bquark}^\ast}}\xspace}
\def\Vtbs  {{\ensuremath{V_{\tquark\bquark}^\ast}}\xspace}

\newcommand{\dms}{{\ensuremath{\Delta m_{\squark}}}\xspace}

\newcommand{\DGs}{{\ensuremath{\Delta\Gamma_{\squark}}}\xspace}

\newcommand{\Gs}{{\ensuremath{\Gamma_{\squark}}}\xspace}

\newcommand{\phis}{{\ensuremath{\phi_{\squark}}}\xspace}
\newcommand{\betas}{{\ensuremath{\beta_{\squark}}}\xspace}

\newcommand{\etag}{{\ensuremath{\varepsilon_{\mathrm{tag}}}}\xspace}

\def\AT#1     {\ensuremath{A_{\mathrm{T}}^{#1}}\xspace}           

\def\C#1      {\ensuremath{\mathcal{C}_{#1}}\xspace}                       
\def\Cp#1     {\ensuremath{\mathcal{C}_{#1}^{'}}\xspace}                    
\def\Ceff#1   {\ensuremath{\mathcal{C}_{#1}^{\mathrm{(eff)}}}\xspace}        
\def\Cpeff#1  {\ensuremath{\mathcal{C}_{#1}^{'\mathrm{(eff)}}}\xspace}       
\def\Ope#1    {\ensuremath{\mathcal{O}_{#1}}\xspace}                       
\def\Opep#1   {\ensuremath{\mathcal{O}_{#1}^{'}}\xspace}                    

\newcommand{\nospaceunit}[1]{\ensuremath{\text{#1}}}       
\newcommand{\aunit}[1]{\ensuremath{\text{\,#1}}}       

\newcommand{\tev}{\aunit{Te\kern -0.1em V}\xspace}
\newcommand{\gev}{\aunit{Ge\kern -0.1em V}\xspace}
\newcommand{\mev}{\aunit{Me\kern -0.1em V}\xspace}
\newcommand{\kev}{\aunit{ke\kern -0.1em V}\xspace}
\newcommand{\ev}{\aunit{e\kern -0.1em V}\xspace}
 
\newcommand{\mevc}{\ensuremath{\aunit{Me\kern -0.1em V\!/}c}\xspace}
\newcommand{\gevc}{\ensuremath{\aunit{Ge\kern -0.1em V\!/}c}\xspace}
\newcommand{\mevcc}{\ensuremath{\aunit{Me\kern -0.1em V\!/}c^2}\xspace}
\newcommand{\gevcc}{\ensuremath{\aunit{Ge\kern -0.1em V\!/}c^2}\xspace}

\def\mum  {\ensuremath{\,\upmu\nospaceunit{m}}\xspace}

\def\fb   {\ensuremath{\aunit{fb}}\xspace}
\def\invfb   {\ensuremath{\fb^{-1}}\xspace}

\def\ps   {\ensuremath{\aunit{ps}}\xspace}
\def\fs   {\aunit{fs}}

\def\invps{\ensuremath{\ps^{-1}}\xspace}

\newcommand{\chisq}{\ensuremath{\chi^2}\xspace}

\newcommand{\chisqip}{\ensuremath{\chi^2_{\text{IP}}}\xspace}

\def\gsim{{~\raise.15em\hbox{$>$}\kern-.85em
          \lower.35em\hbox{$\sim$}~}\xspace}
\def\lsim{{~\raise.15em\hbox{$<$}\kern-.85em
          \lower.35em\hbox{$\sim$}~}\xspace}

\def\sPlot{\mbox{\em sPlot}\xspace}

\def\pt         {\ensuremath{p_{\mathrm{T}}}\xspace}

\def\et         {\ensuremath{E_{\mathrm{T}}}\xspace}

\def\rad{\aunit{rad}}

\def\evtgen     {\mbox{\textsc{EvtGen}}\xspace}

\def\geant      {\mbox{\textsc{Geant4}}\xspace}

\def\photos     {\mbox{\textsc{Photos}}\xspace}

\def\pythia     {\mbox{\textsc{Pythia}}\xspace}

\def\tell1  {TELL1\xspace}
\def\ukl1   {UKL1\xspace}

\usepackage{cite} 
\usepackage{mciteplus}

\usepackage{longtable} 
\usepackage{rotating}
\usepackage{multirow}
\usepackage{hhline}
\usepackage{float}
\usepackage{booktabs}

\begin{document}

\renewcommand{\thefootnote}{\fnsymbol{footnote}}
\setcounter{footnote}{1}


\begin{titlepage}
\pagenumbering{roman}

\vspace*{-1.5cm}
\centerline{\large EUROPEAN ORGANIZATION FOR NUCLEAR RESEARCH (CERN)}
\vspace*{1.5cm}
\noindent
\begin{tabular*}{\linewidth}{lc@{\extracolsep{\fill}}r@{\extracolsep{0pt}}}
\ifthenelse{\boolean{pdflatex}}
{\vspace*{-1.5cm}\mbox{\!\!\!\includegraphics[width=.14\textwidth]{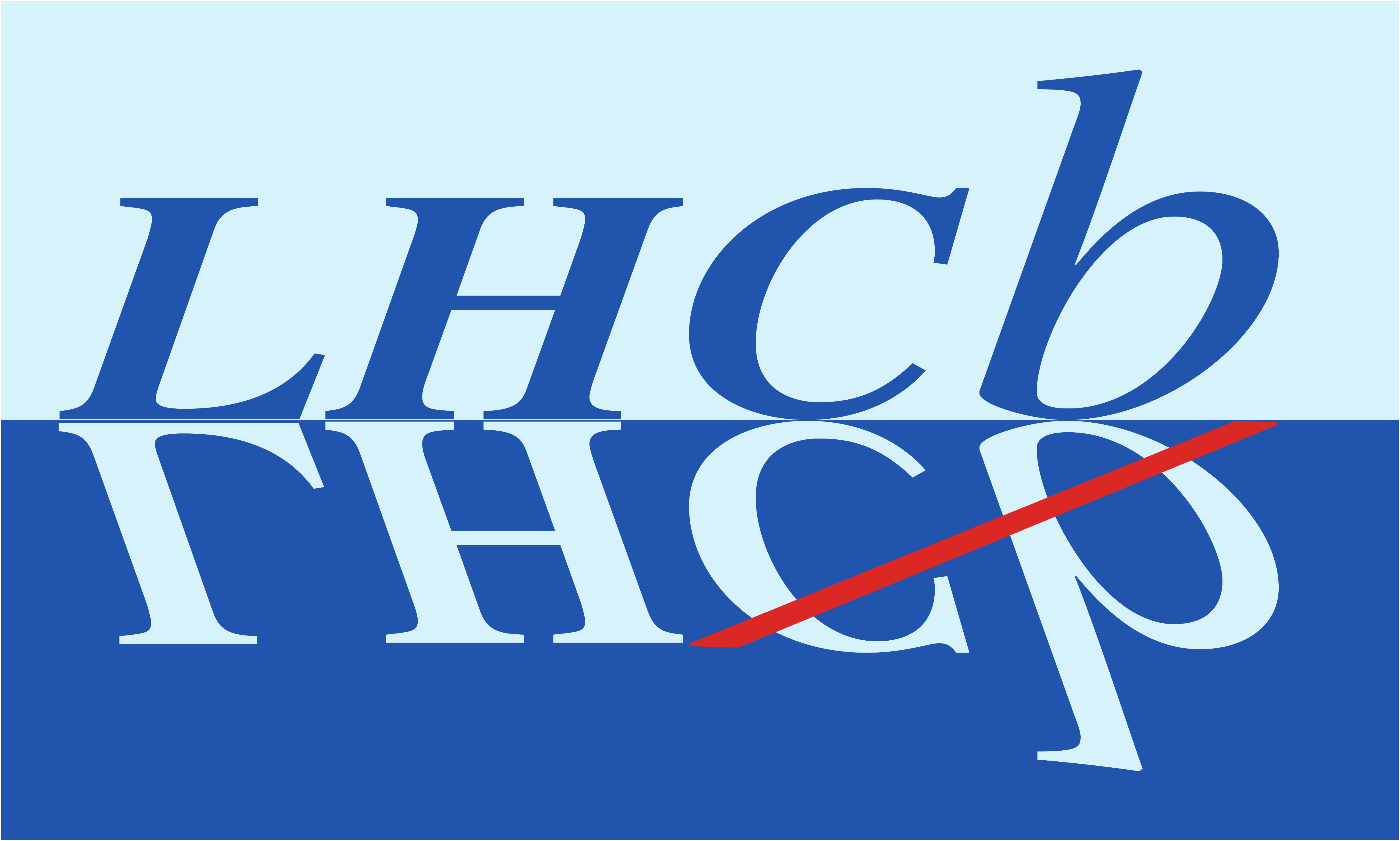}} & &}%
{\vspace*{-1.2cm}\mbox{\!\!\!\includegraphics[width=.12\textwidth]{figs/lhcb-logo.eps}} & &}%
\\
 & & CERN-EP-2021-064 \\  
 & & LHCb-PAPER-2020-042 \\  
 & & 24 November 2021 \\
 & & \\
\end{tabular*}

\vspace*{4.0cm}

{\normalfont\bfseries\boldmath\huge
\begin{center}
  \papertitle 
\end{center}
}

\vspace*{2.0cm}

\begin{center}
\paperauthors\footnote{Authors are listed at the end of this paper.}
\end{center}

\vspace{\fill}

\begin{abstract}
  \noindent
  A flavour-tagged time-dependent angular analysis of \decay{\Bs}{\jpsi\phiz} decays is presented where the \jpsi meson is reconstructed through its decay to an \epem pair. The analysis uses a sample of $pp$ collision data recorded with the \lhcb experiment at centre-of-mass energies of $7$ and $8\tev$, corresponding to an integrated luminosity of $3 \invfb$. The \CP-violating phase and lifetime parameters of the \Bs system are measured to be
  \mbox{$\phis=0.00\pm0.28\pm0.07\rad$}, \mbox{$\DGs=0.115\pm0.045\pm0.011\invps$} and \mbox{$\Gs=0.608\pm0.018\pm0.012\invps$}
  where the first uncertainty is statistical and the second systematic. This is the first time that \CP-violating parameters are measured in the \decay{\Bs}{\jpsi\phiz} decay with an \epem pair in the final state. The results are consistent with previous measurements in other channels and with the Standard Model predictions.
  
\end{abstract}

\vspace*{2.0cm}

\begin{center}
Published in
  Eur.~Phys.~J.~C81 (2021) 1026 
\end{center}

\vspace{\fill}

{\footnotesize 
\centerline{\copyright~\papercopyright. \href{\paperlicenceurl}{\paperlicence}.}}
\vspace*{2mm}

\end{titlepage}


\newpage
\setcounter{page}{2}
\mbox{~}
%
%
%
%


\renewcommand{\thefootnote}{\arabic{footnote}}
\setcounter{footnote}{0}

\cleardoublepage


\pagestyle{plain} 
\setcounter{page}{1}
\pagenumbering{arabic}


\section{Introduction}
\label{sec:Introduction}

The phase difference \phis between direct decays and decays through mixing of \Bs mesons to Charge-Parity (\CP) eigenstates is a \CP-violating observable.
In the Standard Model~(SM), considering \decay{\bquark}{(\cquark\cquarkbar)\squark} transitions and neglecting subleading penguin contributions, this phase is predicted to be $-2\betas$, where $\betas=\arg[-(\Vts\Vtbs)/(\Vcs\Vcbs)]$ and $V_{ij}$ are the elements of the CKM quark-flavour mixing matrix~\cite{Kobayashi:1973fv, *Cabibbo:1963yz}.

The precise measurement of the \phis phase is potentially sensitive to new physics (NP) processes.
The measured phase could be modified if new particles were to contribute to the \Bs--\Bsb mixing amplitudes~\cite{Buras:2009if, *Chiang:2009ev}.
Measurements of \phis using different decay channels with muons in the final state, namely \decay{\Bs}{\jpsi\Kp\Km}~\cite{LHCb-PAPER-2019-013, LHCb-PAPER-2017-008}, \decay{\Bs}{\jpsi\pip\pim}~\cite{LHCb-PAPER-2019-003}, \decay{\Bs}{\psitwos\phiz}~\cite{LHCb-PAPER-2016-027}, and a channel with open charm mesons, \decay{\Bs}{\Dsp\Dsm}~\cite{LHCb-PAPER-2014-051}, have been reported previously by the LHCb collaboration. Measurements of \phis in \decay{\Bs}{\jpsi\phiz} decays with \decay{\jpsi}{\mumu} have also been performed by the \atlas~\cite{Aad:2014cqa, Aad:2020jfw}, \cms \cite{Khachatryan:2015nza}, \cdf \cite{Aaltonen:2012ie} and \dzero \cite{Abazov:2011ry} collaborations. The world-average value of these measurements is \mbox{$\phis=-0.051\pm0.023\rad$}~\cite{PDG2020}. A precise prediction of the \phis phase value is available from global fits of the CKM matrix within the SM. The CKMFitter group result is  \mbox{$\phis=-0.0365^{\,+\,0.0013}_{\,-\,0.0012}\rad$}~\cite{CKMfitter2015} while the UTfit collaboration result is \mbox{$\phis=-0.0370\pm0.0010\rad$}~\cite{UTfit-UT}. 

This paper presents a measurement of \phis using a flavour-tagged time-dependent angular analysis of the \decay{\Bs}{\jpsi\phiz} mode with \decay{\jpsi}{\epem} and \decay{\phiz}{\Kp\Km} decays.\footnote{The inclusion of charge-conjugate processes is implied throughout this paper, unless otherwise noted. For simplicity, the resonance $\phiz(1020)$ is referred to as \phiz here and in the following.} This is the first time that the \decay{\Bs}{\jpsi(}{\epem)\phiz} decay is used to measure \CP-violating observables, and in particular the phase \phis. The analysis is based on a data set corresponding to an integrated luminosity of $3 \invfb$ collected at the $\lhc$ in proton-proton ($pp$) collisions at centre-of-mass energies of $7$ and $8\tev$ by the LHCb experiment. 
The yield of the \decay{\Bs}{\jpsi(}{\epem)\phiz}{(\Kp\Km)} sample amounts to about $10\%$ of that of the previously analysed \decay{\Bs}{\jpsi(}{\mumu)\phiz}{(\Kp\Km)} mode using the same data set~\cite{LHCb-PAPER-2014-059}. The analysis follows closely that of the two muons decay mode, reported in Refs.~\cite{LHCb-PAPER-2019-013, LHCb-PAPER-2019-003}. Relevant changes are described in more detail in this paper. 

A comparison of the two results is of interest given the different main sources of systematic uncertainties induced by the markedly different reconstruction of decays with muons in the final state compared to decays with electrons. These differences arise from the significant bremsstrahlung emission of the electrons and the different signatures exploited in the online trigger selection~\cite{LHCb-PAPER-2017-013,LHCb-PAPER-2019-009,LHCb-PAPER-2019-040}.

The article is structured in the following way. The phenomenological description of the \decay{\Bs}{\jpsi(}{\epem)\phiz}{(\Kp\Km)} decay and the relevant physics observables are described in Sec.~\ref{sec:Phenom}. A brief description of the $\lhcb$ detector, the candidates selection and the background subtraction are outlined in Sec.~\ref{sec:Data}. The relevant inputs to the analysis, namely the resolution, efficiency and the flavour tagging, are detailed in Sec.~\ref{sec:ResEff} and~\ref{sec:FlavTagg}. The maximum-likelihood fit procedure used to determine the physics parameters and the results of the fit are described in Sec.~\ref{sec:Results}, while the evaluation of the systematic uncertainties is discussed in Sec.~\ref{sec:SystUnc}. Finally, conclusions are presented in Sec.~\ref{sec:Concl}.

\section{Phenomenology}
\label{sec:Phenom}

\begin{figure}[t]
  \begin{center}
    \includegraphics[width=0.8\linewidth]{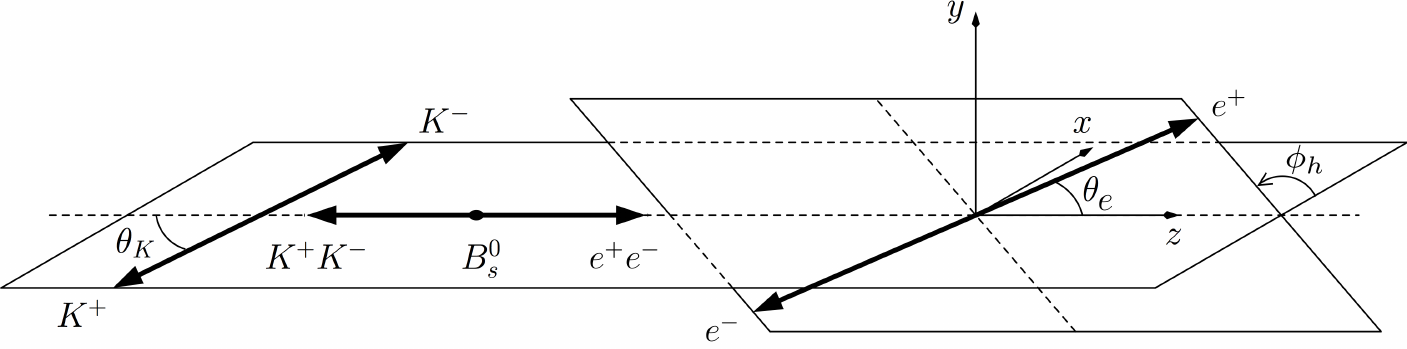}
    \vspace*{-0.5cm}
  \end{center}
  \caption{Definition of the angles in the helicity basis. The polar angle $\theta_K$ $(\theta_e)$ is the angle between the \Kp (\ep) momentum and the direction opposite to the \Bs momentum in the \Kp\Km (\epem) centre-of-mass system, and the $\phi_h$ is the azimuthal angle between the \Kp\Km and \epem decay planes.}
  \label{fig:Phen:Angles}
\end{figure}

The phenomenological aspects of the analysis are presented in Ref.~\cite{LHCb-PAPER-2013-002}. This formalism also holds for the \decay{\Bs}{\jpsi(}{\epem)\phiz}{(\Kp\Km)} decay. 
Angular momentum conservation in the \decay{\Bs}{\jpsi\phiz} decay implies that the final state is an admixture of two \CP-even and one \CP-odd components, with orbital angular momentum of 0 or 2, and 1, respectively. Moreover, along with the three P-wave states of the \decay{\phiz}{\Kp\Km} transition, 
there is also a \CP-odd \Kp\Km component in an S-wave state~\cite{Stone:2008ak}. The \CP-even and \CP-odd components are disentangled by a time-dependent angular analysis, where the angular observables $\Omega=\{\cos\theta_e,\cos\theta_K,\phi_h\}$ are defined in the helicity basis as shown in Fig.~\ref{fig:Phen:Angles}. The polar angle $\theta_K$ $(\theta_e)$ is the angle between the \Kp (\ep) momentum and the direction opposite to the \Bs momentum in the \Kp\Km (\epem) centre-of-mass system. The azimuthal angle between the \Kp\Km and \epem decay planes is $\phi_h$. A definition of the angles in terms of the particles momenta can be found in Ref.~\cite{LHCb-PAPER-2013-002}. 

The differential decay rate for \decay{\Bs}{\jpsi\phiz} decay as a function of the decay time and angles can be expressed as a sum of polarisation amplitudes and their interference terms. Each of these can be factorised into a part dependent on the decay time $t$ and a part dependent on the set of angular variables $\Omega$, as
\begin{equation}\label{eq:decayrate}
 G(t,\Omega)\equiv\frac{\mathrm{d}^{4}\Gamma(\Bs\to\jpsi\phiz)}{\mathrm{d}t\,\mathrm{d}\Omega}\propto\sum^{10}_{k=1} h_k(t)f_k(\Omega).
\end{equation}
The time-dependent functions $h_k(t)$ are given as
\begin{equation}\label{eq:decayrate_time}
 h_k(t|\Bs)= N_k e^{-\Gs t}\left[a_k\cosh\frac{\DGs t}{2}+b_k\sinh\frac{\DGs t}{2}+c_k\cos(\dms t)+d_k\sin(\dms t)\right],
 \end{equation}
 \begin{equation}\label{eq:decayrate_time_Bsb}
 h_k(t|\Bsb)= \bar{N}_k e^{-\Gs t}\left[a_k\cosh\frac{\DGs t}{2}+b_k\sinh\frac{\DGs t}{2}-c_k\cos(\dms t)-d_k\sin(\dms t)\right],
\end{equation}
where $\DGs\equiv\Gamma_{\mathrm{L}}-\Gamma_{\mathrm{H}}$ is the decay width difference between the light and the heavy $\B_s$ mass eigenstates, $\dms\equiv m_{\mathrm{H}}-m_{\mathrm{L}}$ is their mass difference, and $\Gs\equiv(\Gamma_{\mathrm{L}}+\Gamma_{\mathrm{H}})/2$ is their average width. The coefficients $N_k$ ($\bar{N}_k$) and $a_k, b_k, c_k, d_k$ can be expressed in terms of \phis and four complex transversity amplitudes $A_i$ ($\bar{A}_i$) at $t = 0$, as detailed in Table~\ref{tab:AngTimeFunction}. The label $i$ takes the values $\{\perp, \parallel, 0\}$ for the three P-wave amplitudes and S for the S-wave amplitude. The amplitudes are parameterised by $|A_i|e^{i\delta_i}$ with the conventions $\delta_0 = 0$ and $|A_{\perp}|^2 + |A_0|^2 + |A_{\parallel}|^2 = 1$. The S-wave fraction is defined as $F_\mathrm{S} = |A_\mathrm{S}|^2/(|A_\mathrm{S}|^2 + |A_{\perp}|^2 + |A_0|^2 + |A_{\parallel}|^2)$. In contrast to Ref.~\cite{LHCb-PAPER-2019-013}, the S-wave parameters are measured in a single range of $m(\Kp\Km)$ within $\pm30\mevcc$ of the known \phiz mass~\cite{PDG2020}. For a particles produced in a \Bs and \Bsb flavour eigenstates, the coefficients in Eqs.~\eqref{eq:decayrate_time} and \eqref{eq:decayrate_time_Bsb}, respectively are given in Table~\ref{tab:AngTimeFunction} together with the angular functions $f_k(\Omega)$, where the $S$, $D$, $C$ coefficients are defined as
\begin{equation}\label{eq:CPVobserv}
 S=-\frac{2|\lambda|}{1+|\lambda|^2}\sin(\phis),\quad D=-\frac{2|\lambda|}{1+|\lambda|^2}\cos(\phis)\quad\mathrm{and}\quad C=\frac{1-|\lambda|^2}{1+|\lambda|^2}.
\end{equation}
The parameter $\lambda$ is related to \CP violation in the interference between mixing and decay, and is defined by $\lambda=\eta_i(q/p)(\bar{A}_i/A_i)$ where the polarisation states $i$ have the \CP eigenvalue $\eta_i=+1$ for $i\in\{0,\parallel\}$ and $\eta_i=-1$ for $i\in\{\perp,\mathrm{S}\}$. The complex parameters $p$ and $q$ relate the mass eigenstates to the flavour eigenstates, $|B_{\mathrm{L,H}}\rangle=p|\Bs\rangle\pm q|\Bsb\rangle$. The \CP-violating phase is defined by $\phis\equiv-\arg(\lambda)$ and is assumed here to be the same for all polarisation states. The value of $|\lambda|$ equals unity in the absence of \CP violation in decay~\cite{CKMfitter2005, Lenz:2006hd, Artuso:2015swg}. In this paper, the \CP violation in $\B_s$ meson mixing is assumed to be negligible, following the measurements in Refs.~\cite{LHCb-PAPER-2013-033, LHCb-PAPER-2016-013}.

\begin{table}[t]
  \caption{Definition of angular and time-dependent functions for \Bs and \Bsb mesons.}
\begin{center}
\scalebox{0.73}{
\begin{tabular}{cccccccc}
    \toprule
        $k$    & $f_k(\theta_K,\theta_e,\phi_h)$ & $N_k$ & $\bar{N}_k$ & $a_k$ & $b_k$ &  $c_k$ & $d_k$ \\ 
    \midrule
    1   & $2\cos^2\theta_K\sin^2\theta_e$ & $|A_0|^2$ &$|\bar{A}_0|^2$ &  1 & $D$ & $C$ & $-S$\\
    2   & $\sin^2\theta_K(1-\sin^2\theta_e\cos^2\phi_h)$ & $|A_{\parallel}|^2$ & $|\bar{A}_{\parallel}|^2$ & 1 & $D$ & $C$ & $-S$\\
    3   & $\sin^2\theta_K(1-\sin^2\theta_e\sin^2\phi_h)$ & $|A_{\perp}|^2$ &$|\bar{A}_{\perp}|^2$ &  1 & $-D$ & $C$ & $S$\\
    4   & $\sin^2\theta_K\sin^2\theta_e\sin2\phi_h$ & $|A_{\parallel}A_{\perp}|$ & $|\bar{A}_{\parallel}\bar{A}_{\perp}|$ &  $C\sin(\delta_{\perp}-\delta_{\parallel})$ & $S\cos(\delta_{\perp}-\delta_{\parallel})$ & $\sin(\delta_{\perp}-\delta_{\parallel})$ & $D\cos(\delta_{\perp}-\delta_{\parallel})$\\
    5   & $\frac{1}{2}\sqrt{2}\sin2\theta_K\sin2\theta_e\cos\phi_h$ & $|A_0A_{\parallel}|$ & $|\bar{A}_0\bar{A}_{\parallel}|$ & $\cos(\delta_{\parallel}-\delta_0)$  & $D\cos(\delta_{\parallel}-\delta_0)$ & $C\cos(\delta_{\parallel}-\delta_0)$ & $-S\cos(\delta_{\parallel}-\delta_0)$\\
    6   & $-\frac{1}{2}\sqrt{2}\sin2\theta_K\sin2\theta_e\sin\phi_h$ & $|A_0A_{\perp}|$ & $|\bar{A}_0\bar{A}_{\perp}|$ & $C\sin(\delta_{\perp}-\delta_0)$  & $S\cos(\delta_{\perp}-\delta_0)$ & $\sin(\delta_{\perp}-\delta_0)$ & $D\cos(\delta_{\perp}-\delta_0)$\\
    7   & $\frac{2}{3}\sin^2\theta_e$ & $|A_\mathrm{S}|^2$ & $|\bar{A}_\mathrm{S}|^2$ &  1 & $-D$ & $C$ & $S$\\
    8   & $\frac{1}{3}\sqrt{6}\sin\theta_K\sin2\theta_e\cos\phi_h$ & $|A_\mathrm{S}A_{\parallel}|$ & $|\bar{A}_\mathrm{S}\bar{A}_{\parallel}|$ & $C\cos(\delta_{\parallel}-\delta_\mathrm{S})$  & $S\sin(\delta_{\parallel}-\delta_\mathrm{S})$ & $\cos(\delta_{\parallel}-\delta_\mathrm{S})$ & $D\sin(\delta_{\parallel}-\delta_\mathrm{S})$\\
    9   & $-\frac{1}{3}\sqrt{6}\sin\theta_K\sin2\theta_e\sin\phi_h$ & $|A_\mathrm{S}A_{\perp}|$ & $|\bar{A}_\mathrm{S}\bar{A}_{\perp}|$ & $\sin(\delta_{\perp}-\delta_\mathrm{S})$  & $-D\sin(\delta_{\perp}-\delta_\mathrm{S})$ & $C\sin(\delta_{\perp}-\delta_\mathrm{S})$ & $S\sin(\delta_{\perp}-\delta_\mathrm{S})$\\
    10   & $\frac{4}{3}\sqrt{3}\cos\theta_K\sin^2\theta_e$ & $|A_\mathrm{S}A_0|$ & $|\bar{A}_\mathrm{S}\bar{A}_0|$ & $C\cos(\delta_0-\delta_\mathrm{S})$  & $S\sin(\delta_0-\delta_\mathrm{S})$ & $\cos(\delta_0-\delta_\mathrm{S})$ & $D\sin(\delta_0-\delta_\mathrm{S})$\\
    \bottomrule
  \end{tabular}}\end{center}
\label{tab:AngTimeFunction}
\end{table}

\section{Detector, data set and selection}
\label{sec:Data}

The LHCb detector~\cite{LHCb-DP-2008-001, *LHCb-DP-2014-002} is a single-arm forward spectrometer covering the pseudorapidity range \mbox{$2<\eta<5$}, designed for the study of particles containing \bquark or \cquark quarks. The detector includes a high-precision tracking system consisting of a silicon-strip vertex detector surrounding the $pp$ interaction region, a large area silicon-strip detector located upstream of a dipole magnet with a bending power of about 4 Tm, and three stations of silicon-strip detectors and straw drift tubes placed downstream of the magnet. The tracking system provides a measurement of momentum, $p$, of charged particles with a relative uncertainty that varies from $0.5\%$ at low momentum to $1.0\%$ at $200\gevc$. 
The minimum distance of a track to a primary $pp$ collision vertex (PV), the impact parameter (IP), is measured with a resolution of $(15+29/\pt)\mum$, where \pt is the component of the momentum transverse to the beam in \gevc. Different types of charged hadrons are distinguished using information from two ring-imaging Cherenkov detectors (RICH). Photons, electrons, and hadrons are identified by a calorimeter system consisting of scintillating-pad and preshower detectors, an electromagnetic calorimeter (ECAL), and a hadronic calorimeter. 
Muons are identified by a system composed of alternating layers of iron and multiwire proportional chambers.

Samples of simulated events are used to optimise the signal selection, to derive the angular efficiency and to correct the decay-time efficiency. The simulated $pp$ collisions are generated using \pythia~\cite{Sjostrand:2006za, *Sjostrand:2007gs} with a specific \lhcb configuration~\cite{LHCb-PROC-2010-056}. The decays of hadronic particles are described by \evtgen~\cite{Lange:2001uf}, in which final-state radiation is generated using \photos~\cite{davidson2015photos}. The interaction of the generated particles with the detector and its response are implemented using \geant toolkit~\cite{Allison:2006ve, *Agostinelli:2002hh}, as described in Ref.~\cite{LHCb-PROC-2011-006}.

The online candidate selection is performed by a trigger~\cite{LHCb-DP-2012-004}, which consists of a hardware stage, based on information from the calorimeter and muon systems, followed by a software stage, which applies a full decay reconstruction.
At the hardware stage, events are required to have a hadron or electron with a high transverse-energy deposit in the calorimeters, $\et>3\gev$ and $\et>3.68\gev$, respectively.
The subsequent software trigger is implemented as two separate levels that further reduce the event rate.
The first level is designed to select decays which are displaced from all PVs. 
At the second level, \decay{\Bs}{\jpsi\phiz} candidates are selected by identifying events containing a pair of oppositely charged kaons with an invariant mass within $\pm30\mevcc$ of the known \phiz-meson mass~\cite{PDG2020} or by using topological $b$-hadron triggers. These topological triggers require a two-, three- or four-track secondary vertex with a large sum of the \pt of the charged particles and significant displacement from all PVs. 
A multivariate algorithm~\cite{BBDT} is used for the identification of secondary vertices consistent with the decay of a $b$ hadron. The trigger signals are associated with reconstructed particles in the offline selection.
The candidate selection is devised in order to minimise the impact on the decay-time efficiency.

Electrons radiate bremsstrahlung photons when travelling through the detector material.
For events where the photons are emitted upstream of the spectrometer magnet, the photon and the electron deposit their energy in different ECAL cells, and the electron momentum measured by the tracking system is underestimated. Neutral energy deposits in the ECAL compatible with being emitted by the electron are used to correct for this effect. The limitations of the recovery technique degrade the resolution of the reconstructed invariant masses of both the di-electron pair and the \Bs candidate~\cite{LHCb-PAPER-2017-013}.

In the offline selection, \jpsi candidates are formed from two oppositely charged tracks identified as electrons, and \phiz candidates from pairs of oppositely charged tracks identified as kaons. The pairs of tracks need to form a good quality vertex.
The electron candidates are required to have \mbox{$\pt>0.5\gevc$} and di-electron invariant mass \mbox{$m(\epem)\in[2.5,3.3]\gevcc$}, where a wider range compared to the dimuon mode analysis is chosen to account for the radiative tail arising due to bremsstrahlung.
The \pt of the \phiz candidate is required to be larger than $1\gevc$.

The \jpsi and \phiz candidates that are consistent with originating from a common vertex are combined to form \Bs candidates. The mass of the \Bs candidates is required to be in the range \mbox{$m(\epem\Kp\Km)\in[4.7,5.6]\gevcc$}. The reconstructed decay time of the \Bs candidate, $t$, is obtained from a kinematic fit with the \jpsi mass constrained to its known value~\cite{PDG2020} and the \Bs candidate constrained to originate from the associated PV.
Each \Bs candidate is associated with the PV that yields the smallest \chisqip, where \chisqip is defined as the difference in the vertex-fit \chisq of a given PV reconstructed with and without the particle under consideration.
The \Bs candidates are selected if they have decay times in the range \mbox{$0.3<t<14\ps$} and decay-time uncertainty estimates \mbox{$\sigma_t<0.12\ps$}. 
The fraction of events containing more than one \Bs candidate within the \mbox{$m(\epem\Kp\Km)$} range is $2.6\%$. All candidates are retained in the subsequent analysis. The impact of allowing multiple candidates per event is negligible.

The main sources of background are partially reconstructed $b$-hadron decays and combinatorial background. The first of these arises from the $\decay{\Bs}{\chicone(1P)(}{\to\jpsi\g)\phiz}$ and $\decay{\Bs}{\psitwos(}{\to\jpsi~X)\phiz}$ decay.\footnote[2]{The symbol $X$ stands for unreconstructed particles.}
The combinatorial background is due to random combination of tracks in the event that pass the candidate selection.
In addition, possible background contributions to the signal region originate from \decay{\Lb}{\jpsi p\Km} and \decay{\Bd}{\jpsi\Kstar(892)^0} decays, where the proton or the \pim meson from the $\Kstar(892)\to \Kp \pim$ decay is misidentified as a \Kp or \Km meson, respectively.

The combinatorial background is suppressed using a boosted decision tree (BDT)~\cite{Breiman, *AdaBoost} analysis, trained using the TMVA toolkit~\cite{Hocker:2007ht,*TMVA4}.
The BDT discriminant is trained using a signal sample of simulated \decay{\Bs}{\jpsi\phiz} decays, and a sample of background from data.
For the background same-sign combinations of electron and/or kaon pairs are chosen with the same selection criteria as for signal. The simulation is corrected to match the distributions observed in data for variables used in the identification of electrons and kaons. 
The eight variables used for the training of the BDT discriminant are the transverse momenta of the \jpsi and \phiz candidates, the vertex \chisq of the \Bs candidate, the \chisq of the kinematic fit of the \Bs candidate with the \jpsi mass constrained to its known value and the electron and kaon identification probability as provided mainly from the RICH and calorimeter systems. The optimal working point for the BDT discriminant is determined using a figure of merit that optimises the statistical power of the selected data sample for the analysis of \phis by taking the number of signal and background candidates into account~\cite{Xie:2009yz}.

The candidates are rejected if the \Kp candidate can also be identified as a proton by a dedicated neural network~\cite{LHCb-DP-2012-003} to suppress any possible contamination from \decay{\Lb}{\jpsi p\Km} decays.
The remaining misidentified background contribution is estimated using simulated samples and amounts to $1\%$ of the expected signal yield for \Lb decays and is negligible for \Bd decays.

 \begin{figure}[t]
   \begin{center}
   \includegraphics[width=.42\linewidth]{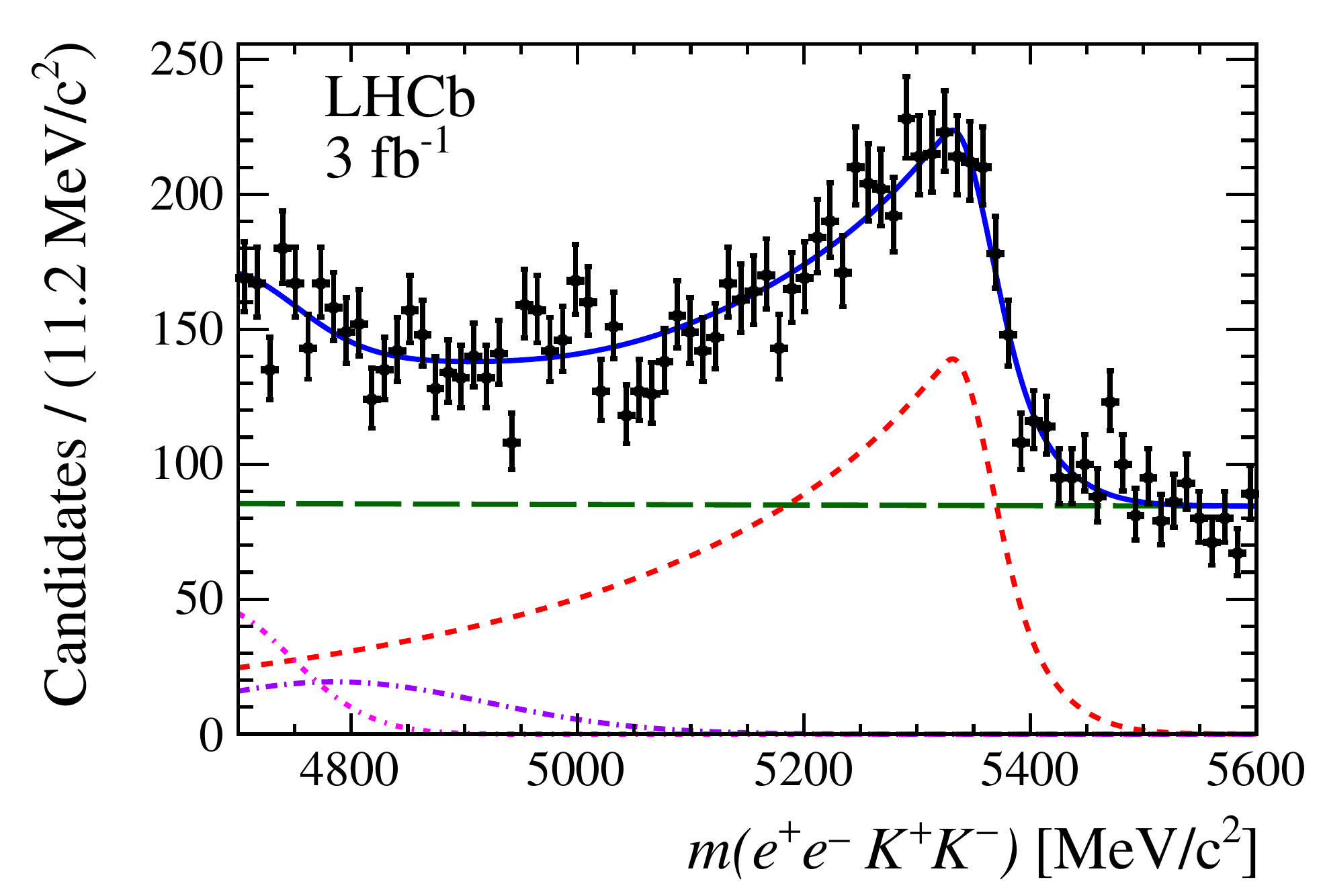}\put(-32,100){\small (a)}\includegraphics[width=.42\linewidth]{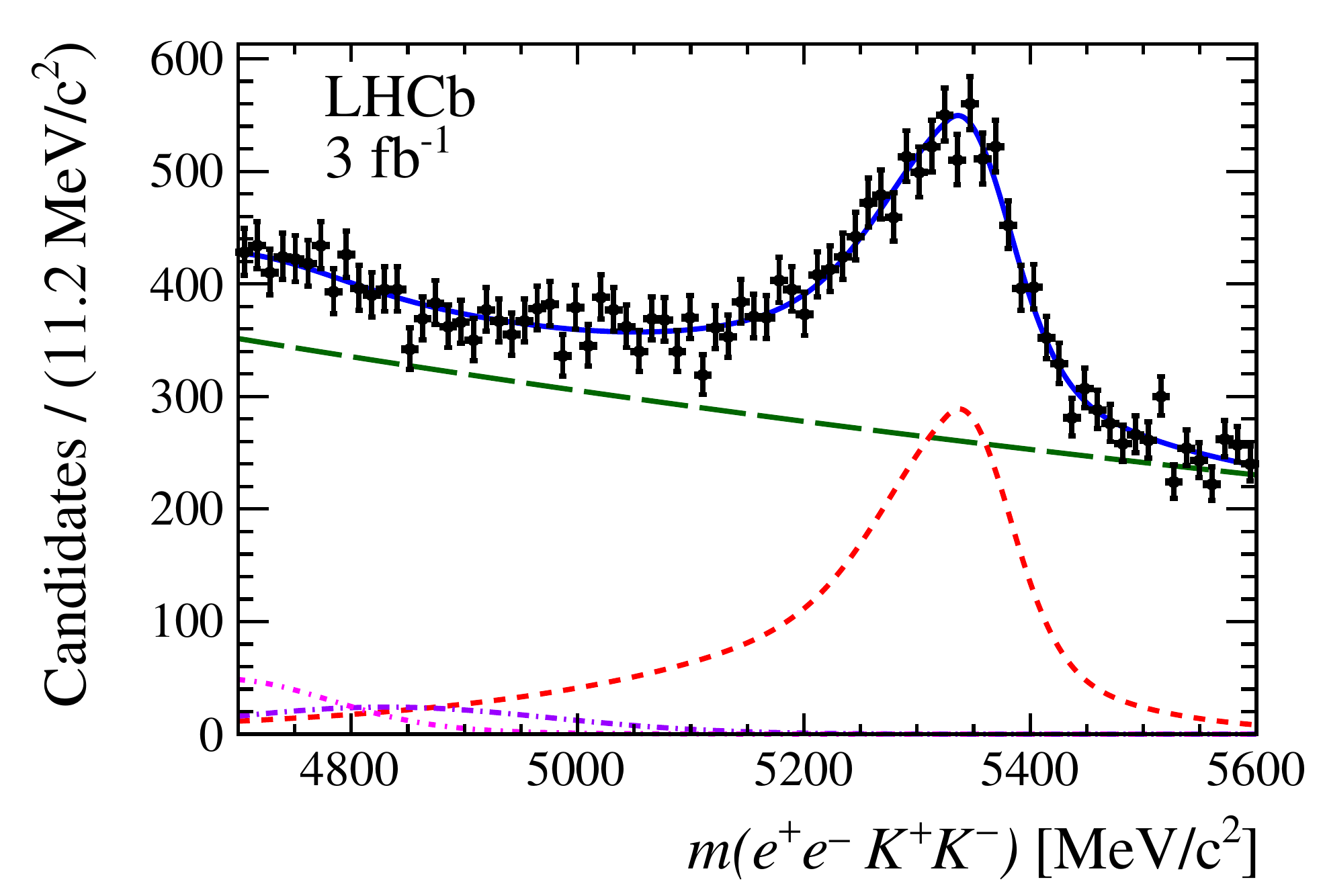}\put(-32,100){\small (b)}\\\includegraphics[width=.42\linewidth]{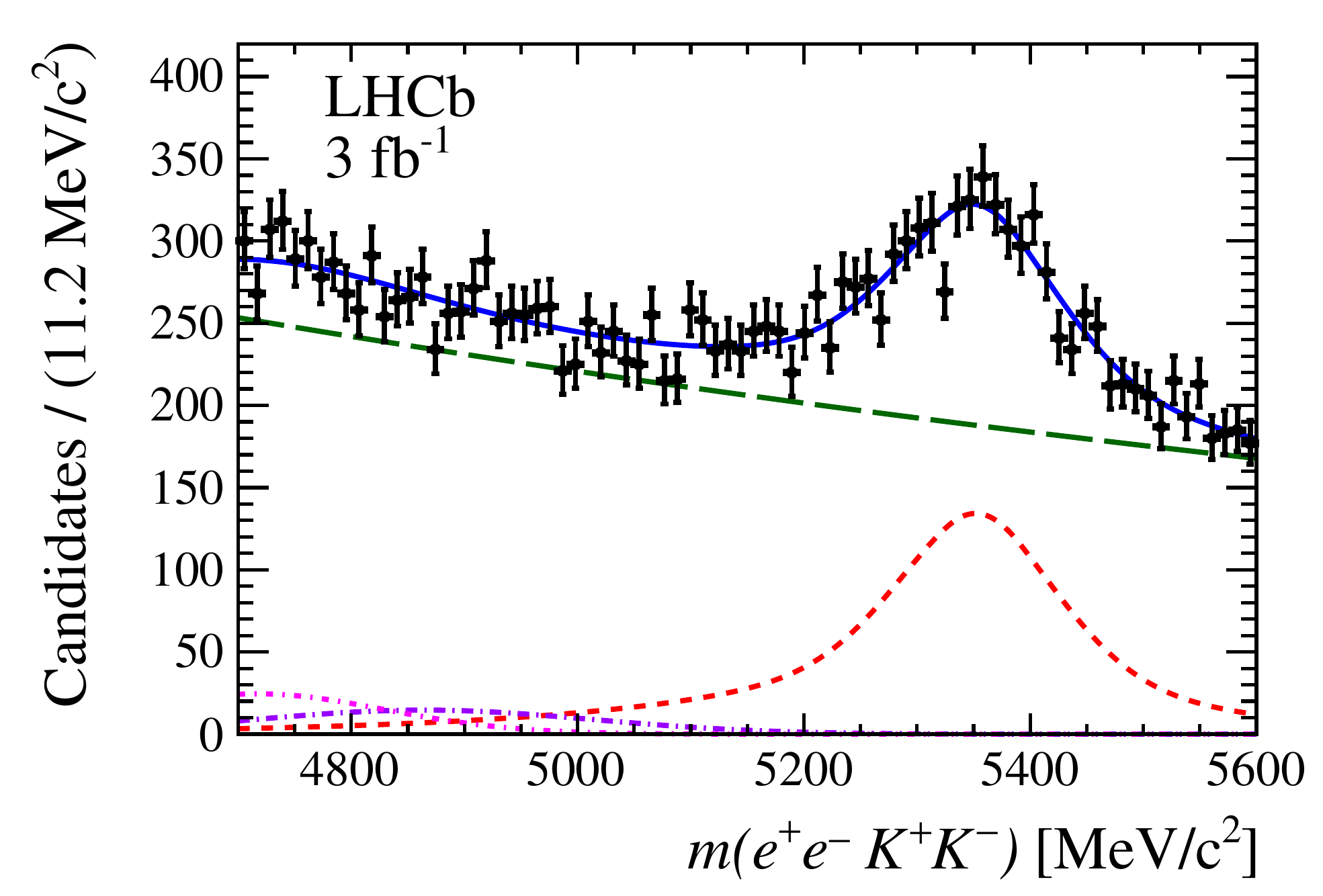}\put(-32,100){\small (c)}
   \includegraphics[width=.42\linewidth]{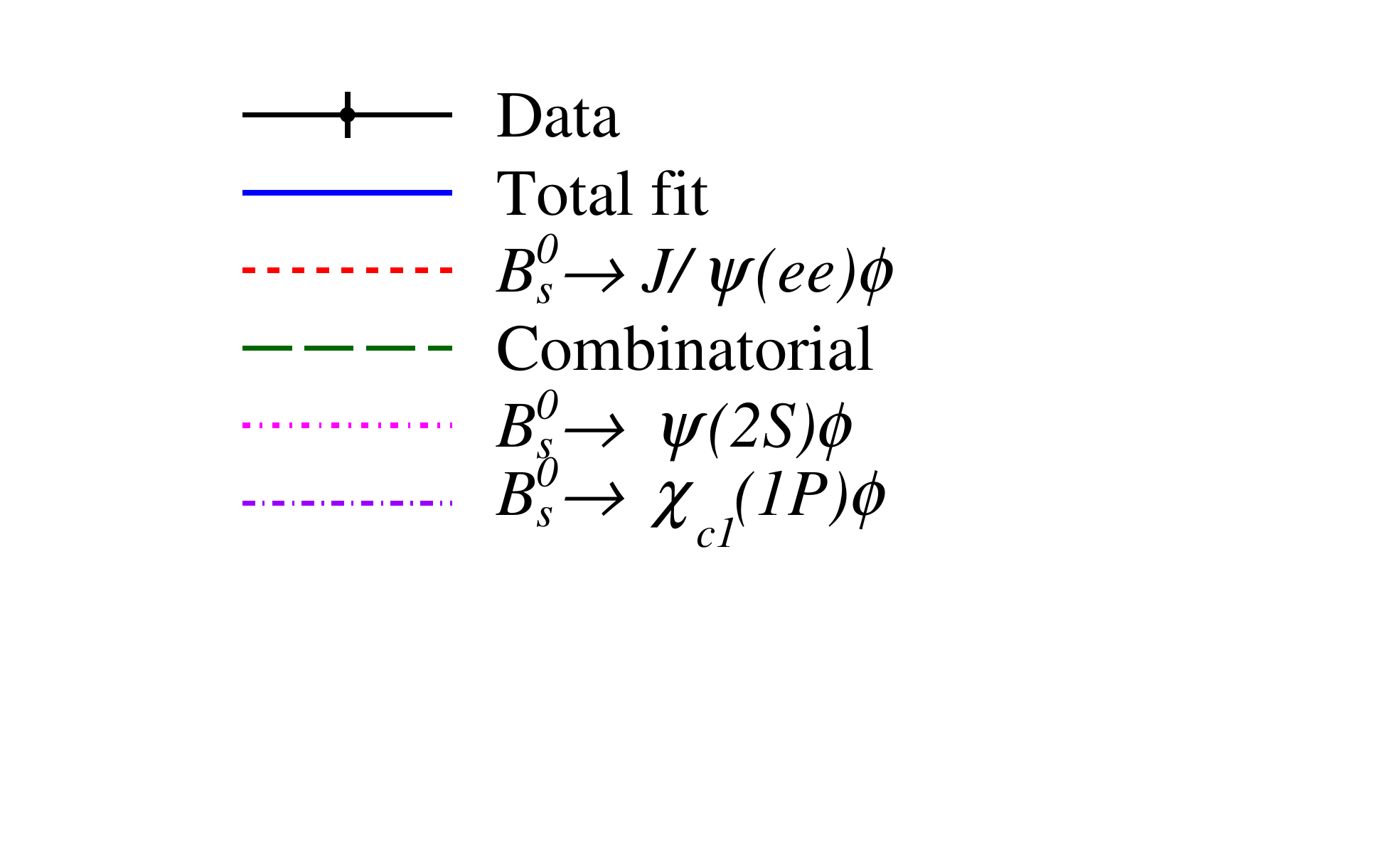}\\
   \vspace*{-0.5cm}
   \end{center}
   \caption{Distribution of $m(\epem\Kp\Km)$ for selected \decay{\Bs}{\jpsi\phiz} candidates divided into three categories: (a) zero, (b) one and (c) both electrons with bremsstrahlung correction. The blue solid line shows the total fit which is composed of (red short-dashed line) the signal and the background contributions. The combinatorial background is indicated by the green long-dashed line while the partially reconstructed background from the \decay{\Bs}{\psitwos\phiz} and \decay{\Bs}{\chicone(1P)\phiz} decays are indicated by pink and purple dash-dotted lines, respectively.}
   \label{fig:Bsmass}
 \end{figure}

Figure~\ref{fig:Bsmass} shows the distribution of $m(\epem\Kp\Km)$ for the selected \decay{\Bs}{\jpsi\phiz} candidates. In order to describe better the left tail of the $m(\epem\Kp\Km)$ distribution, the sample is split into three categories by the number of electron candidates: zero, one or both electrons of the pair that received bremsstrahlung corrections. 
An extended maximum-likelihood fit is made to the unbinned $m(\epem\Kp\Km)$ distribution.

In the fit the signal component is described by the sum of two Crystal Ball (CB) functions~\cite{Skwarnicki:1986xj} and the combinatorial background by an exponential function.
The partially reconstructed background components from \decay{\Bs}{\chicone(1P)\phiz} and \decay{\Bs}{\psitwos\phiz} decays are modelled using a Gaussian function and the sum of two Gaussian functions, respectively.
The parameters that describe the shape of the signal candidates and the partially reconstructed background are fixed to values obtained from simulation.
The core widths and the common mean of the CB functions are left free in the fit. 
The fit to the three categories gives a yield of $(1.27\pm0.05)\times10^4$ signal candidates where the uncertainty is statistical only.

The fit results are used to assign per-candidate weights via the \sPlot technique with $m(\epem\Kp\Km)$ as the discriminating variable~\cite{Pivk:2004ty}. This is used to subtract the background contribution in the maximum-likelihood fit described in Sec.~\ref{sec:Results}.
As the three categories are statistically independent further steps of the analysis are performed on the combined sample.

\section{Detector resolution and efficiency}
\label{sec:ResEff}  

The finite decay-time resolution is a diluting factor that
will affect the relative precision of $\phis$ and has to be accounted for.
The way this is introduced into the analysis is described in Sec.~\ref{sec:Results}.
The assumed decay-time resolution model, $\mathcal{R}$, consists of a sum of two Gaussian distributions with their widths depending on the per-candidate decay-time uncertainty determined by the vertex fit as detailed in Ref.~\cite{LHCb-PAPER-2014-059}.
The parameters of this model are loosely constrained in the fit of the \decay{\Bs}{\jpsi(}{\epem)\Kp\Km} decay to the values determined using an identical model from a sample of \decay{\jpsi}{\mumu} candidates produced at the PV.
They are allowed to vary within a Gaussian constraint of twice the difference of their values between the electron and muon modes as extracted from simulation.
The loose constraint was selected to minimise reliance of the analysis on simulations, increasing further the allowed variation does not impact the results.
The parameters are determined from the unbinned maximum-likelihood fit, as described in Sec.~\ref{sec:Results}.
Taking into account the $\sigma_t$ distribution of the \Bs signal, the resulting effective resolution is $45.6\pm0.5\fs$.

\begin{figure}[t]
   \begin{center}
   \includegraphics[width=.42\linewidth]{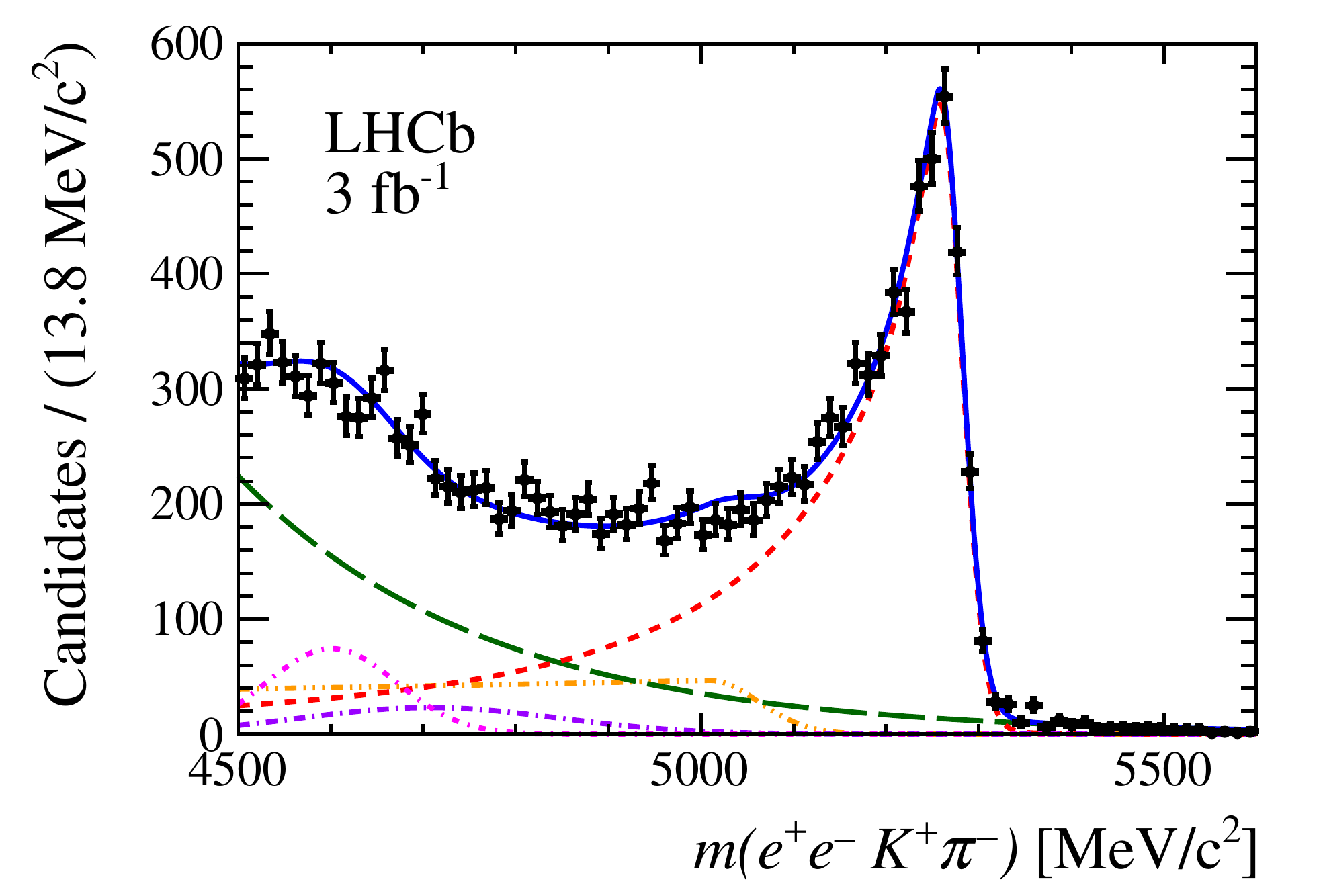}\put(-32,100){\small (a)}
   \includegraphics[width=.42\linewidth]{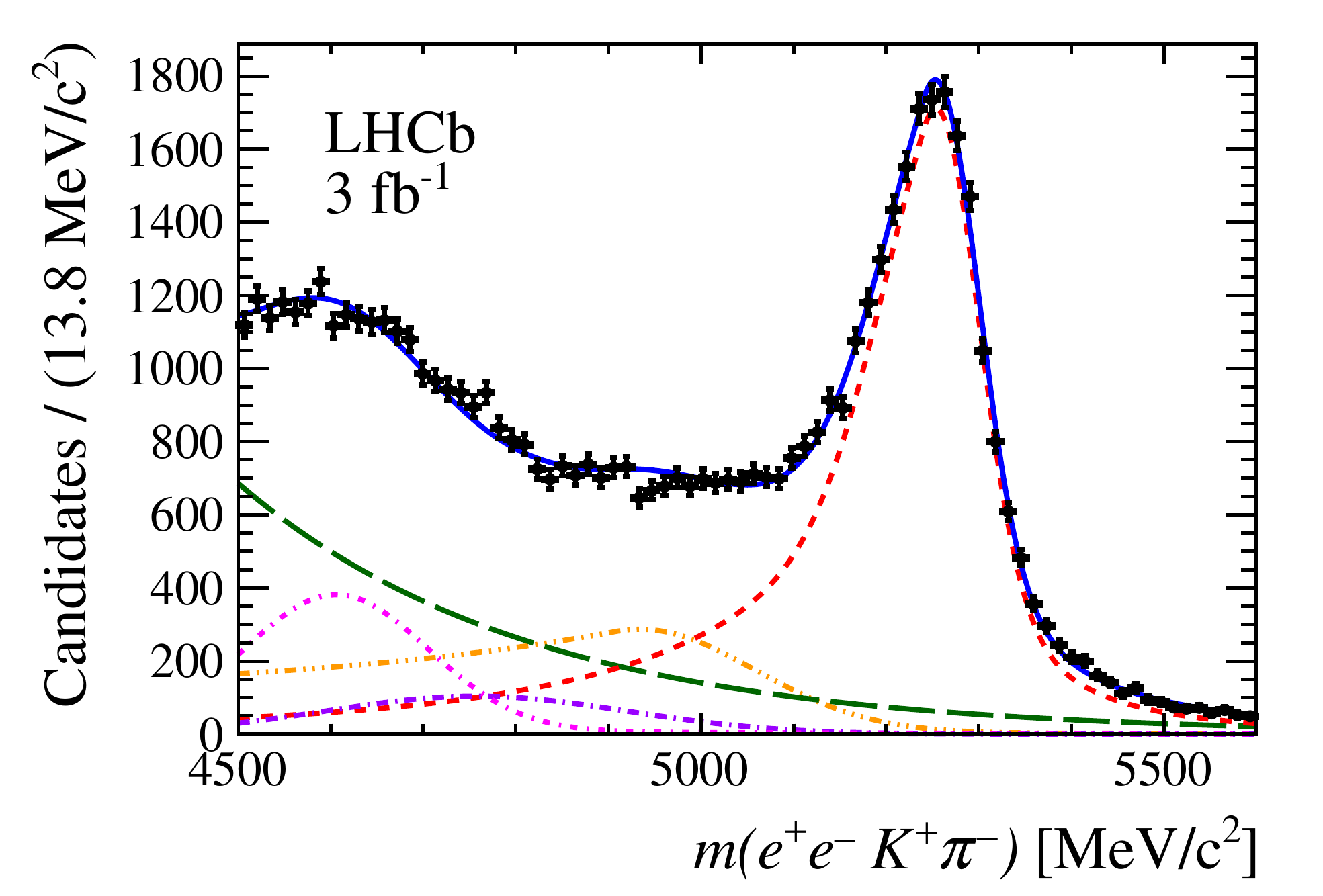}\put(-32,100){\small (b)}\\\includegraphics[width=.42\linewidth]{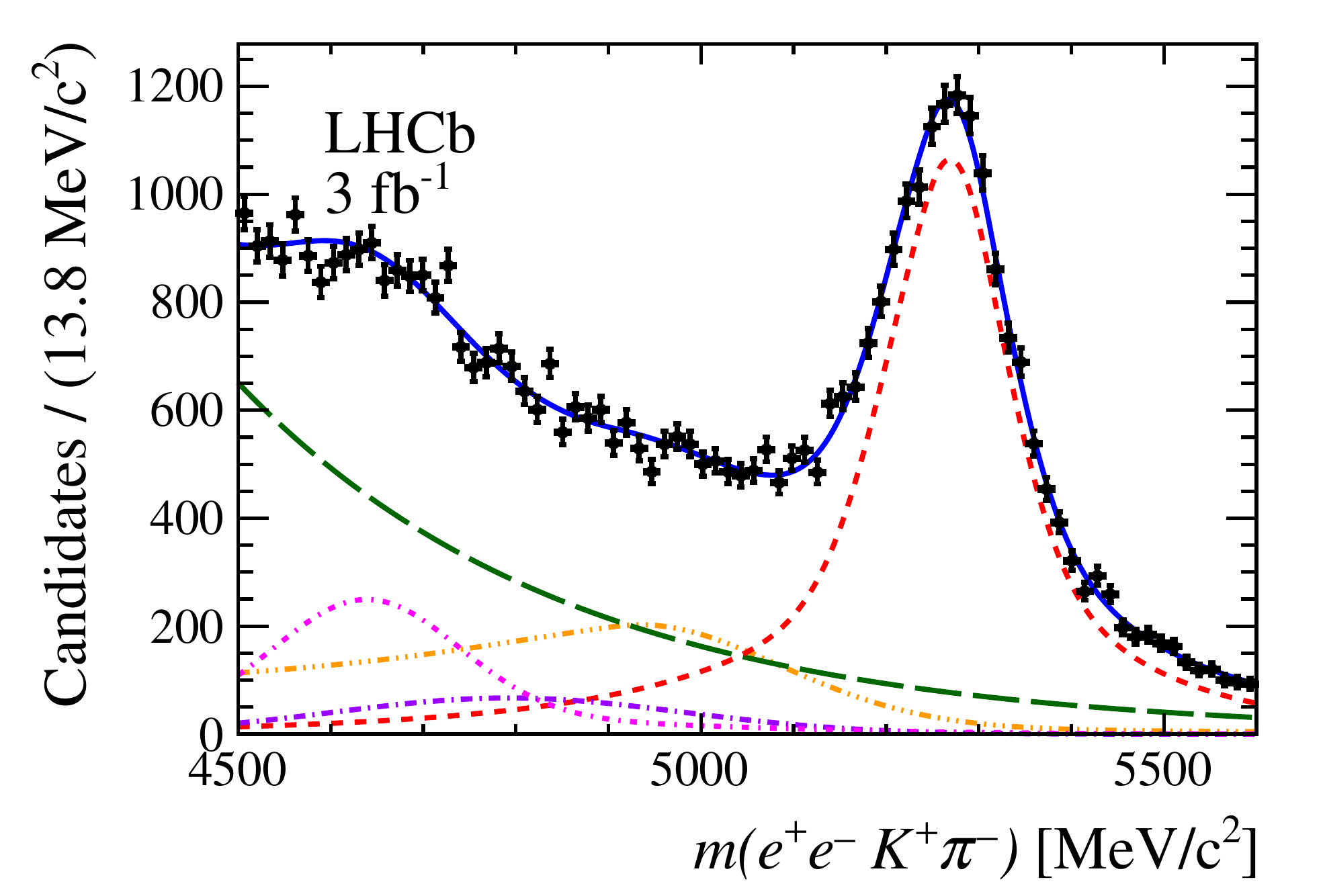}\put(-32,100){\small (c)}
   \includegraphics[width=.42\linewidth]{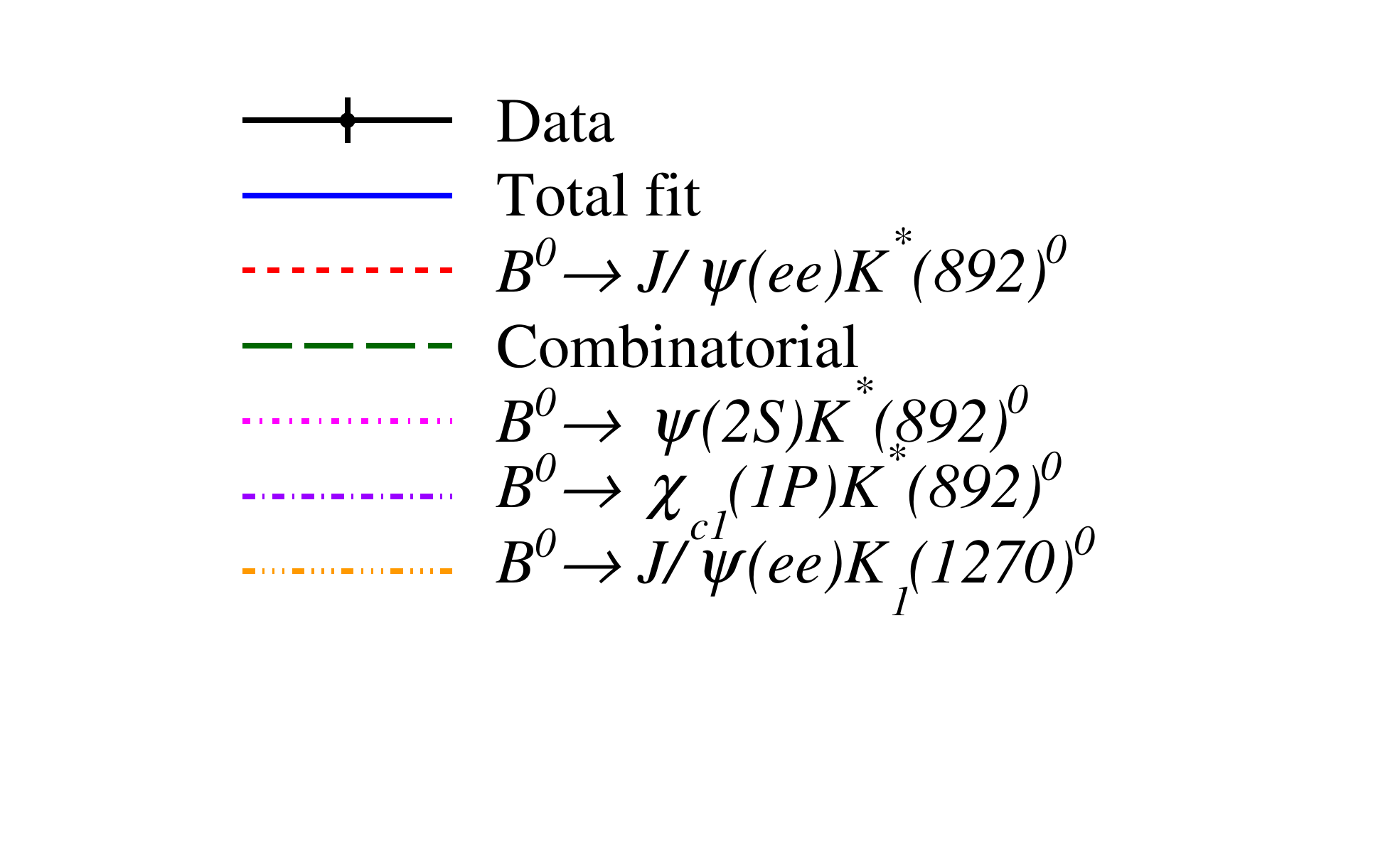}\\
   \vspace*{-0.5cm}
   \end{center}
   \caption{Distribution of $m(\epem\Kp\pim)$ for selected \decay{\Bd}{\jpsi\Kstar(892)^0} candidates divided into three categories: (a) zero, (b) one and (c) both electrons with bremsstrahlung correction. The blue solid line shows the total fit which is composed of (red short-dashed line) the signal and the background contributions. The combinatorial background is indicated by the green long-dashed line while the partially reconstructed background from the \decay{\Bd}{\psitwos\Kstar(892)^0}, \decay{\Bd}{\chicone(1P)\Kstar(892)^0} and \decay{\Bd}{\jpsi K_1(1270)^0} decays are indicated by pink, purple and yellow dash-dotted lines, respectively.}
   \label{fig:Bdmass}
 \end{figure}

Due to the displacement requirements made on signal tracks in the trigger and offline selections, the reconstruction efficiency depends on the decay time of the \Bs candidate. The efficiency is determined with the same method as described in Ref.~\cite{LHCb-PAPER-2016-027}, by using the control channel \decay{\Bd}{\jpsi\Kstar(892)^0}, with \decay{\jpsi}{\epem} and \decay{\Kstar(892)^0}{\Kp\pim} decays. 

The decay-time dependence of the signal efficiency is determined as

\begin{equation}
 \eps^{\Bs}_{\mathrm{data}}(t) = \eps^{\Bd}_{\mathrm{data}}(t)\times\frac{\eps^{\Bs}_{\mathrm{sim}}(t)}{\eps^{\Bd}_{\mathrm{sim}}(t)},
\end{equation}
where $\eps^{\Bd}_{\mathrm{data}}(t)$ is the efficiency of the control channel, determined on data, and $\eps^{\Bs}_{\mathrm{sim}}(t)/\eps^{\Bd}_{\mathrm{sim}}(t)$ is the ratio of efficiencies of the simulated signal and control modes after the selection. The efficiencies are extracted by normalisation to the known lifetimes of \mbox{$\tau_{\Bs}=1.527\pm0.011\ps$} and \mbox{$\tau_{\Bd}=1.520\pm0.004\ps$}~\cite{PDG2020}. The second term accounts for the small differences in the decay time and kinematics between the signal and the control modes.
The control channel efficiency is defined as $\eps^{\Bd}_{\mathrm{data}}(t)=N^{\Bd}_{\mathrm{data}}(t)/N^{\Bd}_{\mathrm{gen}}(t)$ where $N^{\Bd}_{\mathrm{data}}(t)$ is the number of the \decay{\Bd}{\jpsi\Kstar(892)^0} decays in a given time bin as determined using \sPlot technique~\cite{Pivk:2004ty} with $m(\epem\Kp\pim)$ as discriminating variable. The $N^{\Bd}_{\mathrm{gen}}(t)$ is the number of events generated from an exponential distribution with lifetime $\tau_{\Bd}$~\cite{PDG2020}. The analysis is not sensitive to the absolute scale of the efficiency.

The \decay{\Bd}{\jpsi\Kstar(892)^0} decay is selected using trigger, selection and BDT requirements similar to those used for the signal, adapted to the different final states. The background contribution to the control sample from the misidentification of final-state particles from the \decay{\Lb}{\jpsi p\pim} decay is estimated to be $0.06\%$ of the expected signal yield, while the background contribution from \decay{\Bs}{\jpsi\phiz} decays is negligible.

The $m(\epem\Kp\pim)$ invariant-mass distribution is shown in Fig.~\ref{fig:Bdmass} divided into the three bremsstrahlung categories, as for the signal sample. The contribution from \mbox{\decay{\Bd}{\jpsi\Kstar(892)^0}} decays is described by the sum of two CB functions while an exponential function is used to describe the combinatorial background. Similarly to the signal sample, partially reconstructed background arises from \Bd decays where one or more particles are not reconstructed; background components stemming from
$\decay{\Bd}{\chicone(1P)(}{\to\jpsi\g)\Kstar(892)^0}$, $\decay{\Bd}{\psitwos(}{\to\jpsi~X)\Kstar(892)^0}$ and 
$\decay{\Bd}{\jpsi K_1(1270)^0(}{\to\Kstar(892)^0\piz)}$ decays\footnotemark[2] are described using a single Gaussian function, the sum of two Gaussian functions and the sum of two CB functions, respectively. The \decay{\Bd}{\jpsi\Kstar(892)^0} yield is found to be $(5.45\pm0.05)\times10^4$ signal candidates.

\begin{figure}[t]
 \begin{center}
  \includegraphics[width=0.4\linewidth]{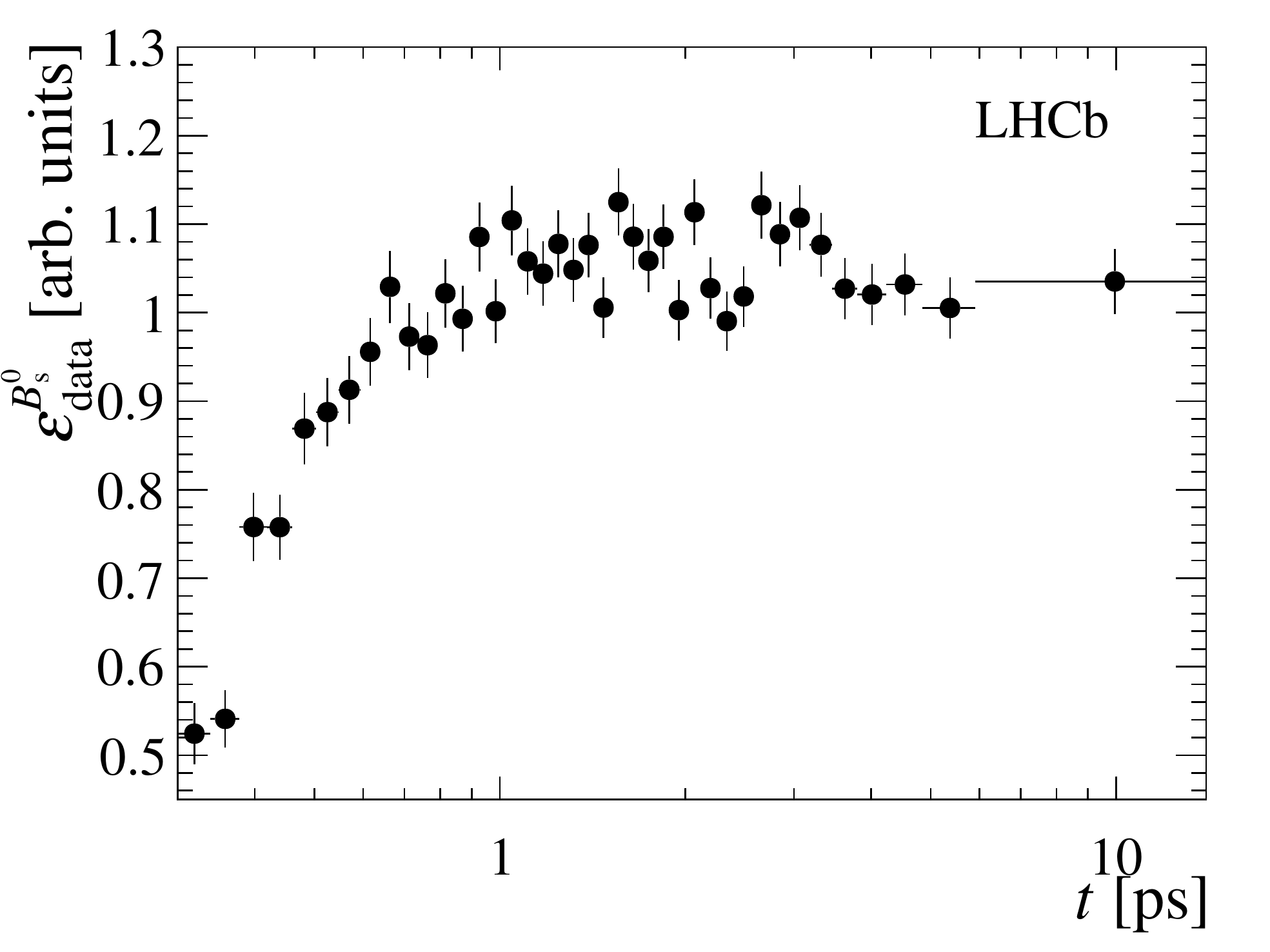}
  \vspace*{-0.5cm}
    \end{center}
    \caption{Signal efficiency as a function of the decay time, $\eps^{\Bs}_{\mathrm{data}}(t)$, scaled by the average efficiency.}
   \label{fig:TimeAccBs}
 \end{figure}

The decay-time efficiency for the \decay{\Bs}{\jpsi\phiz} signal is shown in Fig.~\ref{fig:TimeAccBs}. The efficiency is relatively uniform at high values of decay time but decreases at low decay times due to the selection criteria that require displaced tracks.

\begin{figure}[t]
 \begin{center}
  \includegraphics[width=0.95\linewidth]{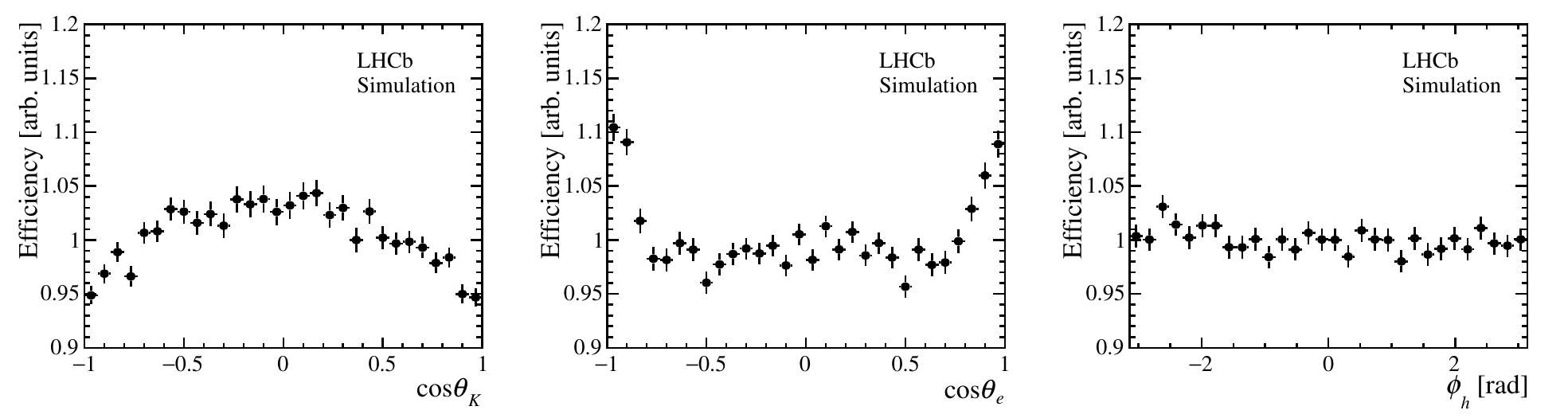}
  \vspace*{-0.5cm}
    \end{center}
    \caption{Efficiency projected onto (left) $\cos\theta_K$, (middle) $\cos\theta_e$ and (right) $\phi_h$ obtained from a simulated \decay{\Bs}{\jpsi\phiz} sample, scaled by the average efficiency.}
   \label{fig:AngAccBs}
 \end{figure}

The efficiency as a function of the \decay{\Bs}{\jpsi\phiz} helicity angles is not uniform due to the forward geometry of the LHCb detector and the requirements imposed on the final-state particle momenta.
Projections of the three-dimensional efficiency, $\eps(\Omega)$,  to the three helicity angles are shown in Fig.~\ref{fig:AngAccBs}.
The angular efficiency correction is introduced in the analysis through normalisation integrals in the probability density function describing the signal decays in the fit described in  Sec.~\ref{sec:Results}. The integrals given in Table~\ref{tab:AngIntegral} are calculated using simulated candidates that are subject to the same trigger and selection criteria as the data, following the same technique as in Ref.~\cite{LHCb-PAPER-2013-002}.
The relative efficiency is constant for the azimuthal angle $\phi_h$. A dependence of up to $15\%$ is observed for $\cos\theta_e$ and $\cos\theta_K$. The finite angular resolution has small impact on the results of the analysis and is neglected. A systematic uncertainty is assigned to account for this effect.

\begin{table}[t]
  \caption{Angular acceptance integrals for the simulated sample. The $I_k$ integrals are normalised with respect to the $I_0$ integral.}
\begin{center}\vspace*{-0.5cm}\begin{tabular}{llr}
    \toprule
\multicolumn{2}{c}{$k$} & \multicolumn{1}{c}{$I_k/I_0$} \\
\midrule
  $1$ & $(00)$ & $0.9801\pm0.0014$ \\
  $2$ & $(\parallel\parallel)$ & $1.0200\pm0.0017$\\
  $3$ & $(\perp\perp)$ & $1.0209\pm0.0016$\\
  $4$ & $(\parallel\perp)$ & $0.0003\pm0.0018$\\
  $5$ & $(0\parallel)$ & $0.0008\pm0.0012$\\
  $6$ & $(0\perp)$ & $0.0015\pm0.0012$\\
  $7$ & (SS) & $0.9983\pm0.0011$\\
  $8$ & (S$\parallel$) & $0.0004\pm0.0016$\\
  $9$ & (S$\perp$) & $0.0012\pm0.0016$\\
  $10$ & (S$0$) & $-0.0067\pm0.0036$\\
    \bottomrule
  \end{tabular}\end{center}
\label{tab:AngIntegral}
\end{table}

\section{Flavour tagging}
\label{sec:FlavTagg}

The $\B_s$ candidate flavour at production is determined by two independent categories of flavour tagging algorithms, the opposite-side (OS) taggers~\cite{LHCb-PAPER-2011-027} and the same-side kaon (SSK) tagger~\cite{LHCb-PAPER-2015-056}, which exploit specific features of the production of \bquark\bquarkbar quark pairs in $pp$ collisions, and their subsequent hadronisation. Each tagging algorithm assigns a tag decision and a mistag probability. The tag decision, $\mathfrak{q}$, takes values $+1$, $-1$, or $0$, if the signal candidate is tagged as \Bs, \Bsb, or is untagged, respectively. The fraction of events in the sample with a nonzero tagging decision gives the efficiency of the tagger, $\varepsilon_{\mathrm{tag}}$. The mistag probability, $\eta$, is estimated event-by-event, and represents the probability that the algorithm assigns a wrong tag decision. It is calibrated using data samples of two flavour specific decays, \decay{\Bpm}{\jpsi(}{\epem)\Kpm} for the OS taggers and \decay{\Bs}{\Dsm\pip} for the SSK tagger, resulting in a corrected mistag probability, $\omega$ $(\bar{\omega})$, for a candidate with initial flavour \Bs (\Bsb).
In case of the SSK algorithm, the calibrated sample of \decay{\Bs}{\Dsm\pip} decays is weighted to match the kinematics of the \decay{\Bs}{\jpsi\phiz} signal decays.
A linear relationship between $\eta$ and $\omega$ is used for the calibration.
The effective tagging power is given by $\etag(1-2\omega)^2$ and for the combined taggers in the \decay{\Bs}{\jpsi(}{\epem)\phiz} signal sample a value of $(5.07\pm0.16)\%$ is obtained.

\section{Maximum-likelihood fit and results}
\label{sec:Results}

The \CP observables are determined by an unbinned maximum-likelihood fit to the background-subtracted candidates in four-dimensions, namely the \Bs decay time and the three helicity angles, with a probability density function (PDF) describing \decay{\Bs}{\jpsi(}{\epem)\phiz} signal decay. 
The negative log-likelihood function to be minimised is given by
\begin{equation}\label{eq:MaxLikFit}
 -\ln\mathcal{L} = -\alpha\sum_{i=1}^{\mathrm{N}} w_i\ln\mathcal{P},
\end{equation}
where N is the total number of candidates. The $w_i$ coefficients are the \sPlot weights~\cite{Pivk:2004ty} computed using $m(\epem\Kp\Km)$ as discriminating variable, and the factor $\alpha=\sum w_i/\sum w^2_i$ is used to account for the correct signal yield in the sample. The PDF, $\mathcal{P}=\mathcal{S}/\int\mathcal{S}\mathrm{d}t\,\mathrm{d}\Omega$, is normalised over the four-dimensional space where 
\begin{equation}\label{eq:SigPDF}
 \mathcal{S}(t,\Omega,\mathfrak{q}^{\mathrm{OS}},\mathfrak{q}^{\mathrm{SSK}}|\eta^{\mathrm{OS}},\eta^{\mathrm{SSK}}) = \mathcal{T}(t',\Omega,\mathfrak{q}^{\mathrm{OS}},\mathfrak{q}^{\mathrm{SSK}}|\eta^{\mathrm{OS}},\eta^{\mathrm{SSK}})\otimes\mathcal{R}(t-t'|\sigma_t)\times\varepsilon^{\Bs}_{\mathrm{data}}(t),
\end{equation}
with the decay-time resolution function, $\mathcal{R}$, defined in Sec.~\ref{sec:ResEff} and
\begin{equation}\label{eq:SigPDF_FT}
\begin{aligned}
  \mathcal{T}(t',\Omega,\mathfrak{q}^{\mathrm{OS}},\mathfrak{q}^{\mathrm{SSK}}|\eta^{\mathrm{OS}},\eta^{\mathrm{SSK}}) & = \left(1+\mathfrak{q}^{\mathrm{OS}}(1-2\omega^{\mathrm{OS}})\right)\left(1+\mathfrak{q}^{\mathrm{SSK}}(1-2\omega^{\mathrm{SSK}})\right) G(t,\Omega)\\
  & + \left(1-\mathfrak{q}^{\mathrm{OS}}(1-2\bar{\omega}^{\mathrm{OS}})\right)\left(1-\mathfrak{q}^{\mathrm{SSK}}(1-2\bar{\omega}^{\mathrm{SSK}})\right) \bar{G}(t,\Omega),
\end{aligned}
\end{equation}
which allows for the inclusion of the information from both tagging algorithms in the computation of the decay rate. The function $G(t,\Omega)$ is defined in Eq.~\eqref{eq:decayrate} and $\bar{G}(t,\Omega)$ is the corresponding function for \Bsb decays. The angular efficiency is included in the normalisation of the PDF via the ten integrals, $I_k=\int\mathrm{d}\Omega\,\varepsilon(\Omega)f_k(\Omega)$. The integrals are pre-calculated using simulation as described in Sec.~\ref{sec:ResEff}.

When using weights from the \sPlot method, the standard uncertainty estimate based on the Hessian matrix will generally not give asymptotically correct confidence intervals~\cite{Langenbruch:2019nwe}.
A bootstrap method~\cite{efron:1979} is used to obtain a correct estimate of the statistical uncertainty.
The weights are recalculated for each bootstrap sample.
In the fit, Gaussian constraints are included for certain nuisance parameters, namely the mixing frequency \mbox{$\dms=17.757\pm0.021\invps$}~\cite{PDG2020}, the tagging calibration parameters, and the time resolution parameters.
The fitting procedure is validated using pseudoexperiments and simulated \mbox{$\decay{\Bs}{\jpsi(}{\epem)\phiz}$} decays. 

\begin{figure}[t]
  \begin{center}
  \includegraphics[width=0.4\linewidth]{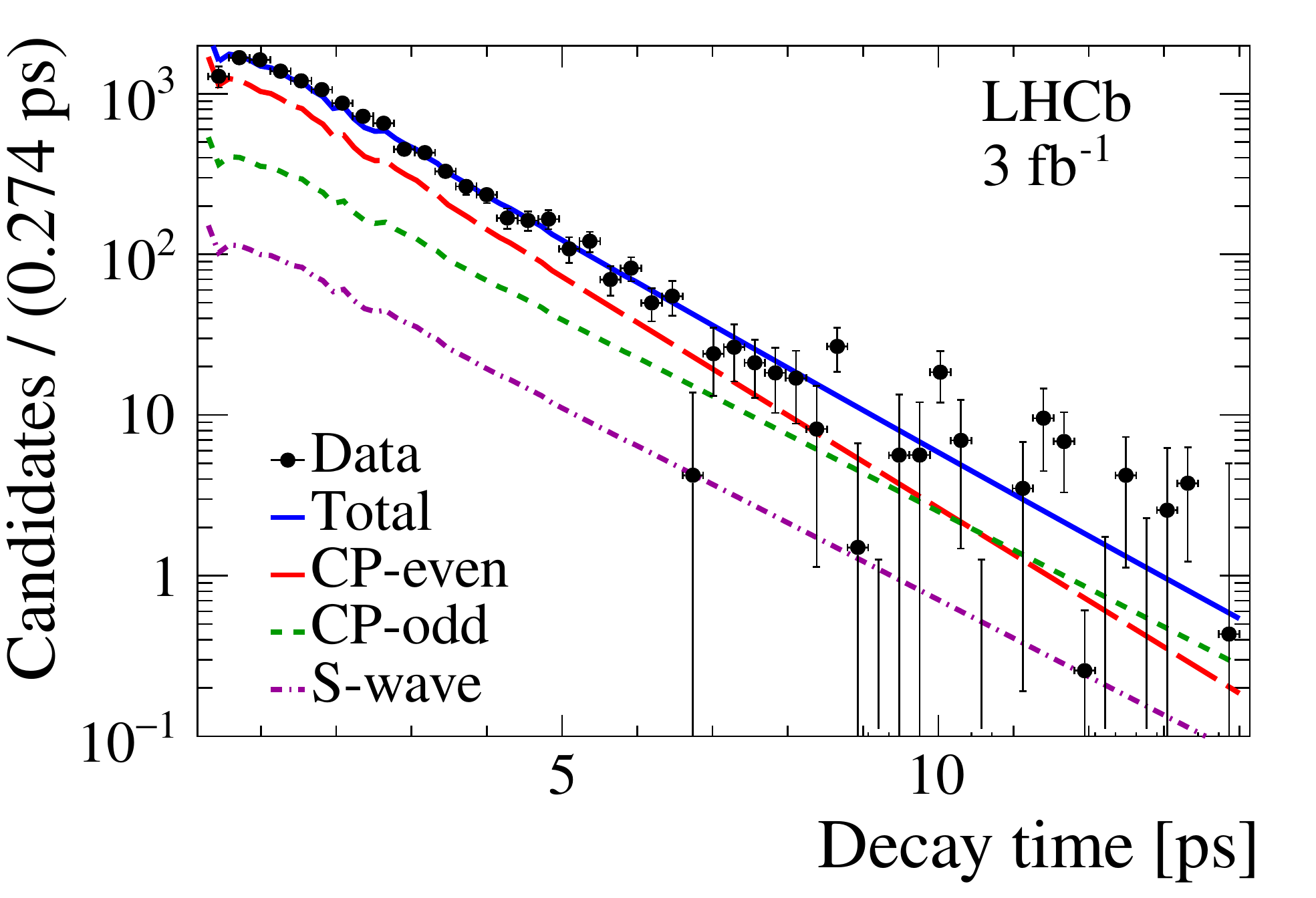}\includegraphics[width=0.4\linewidth]{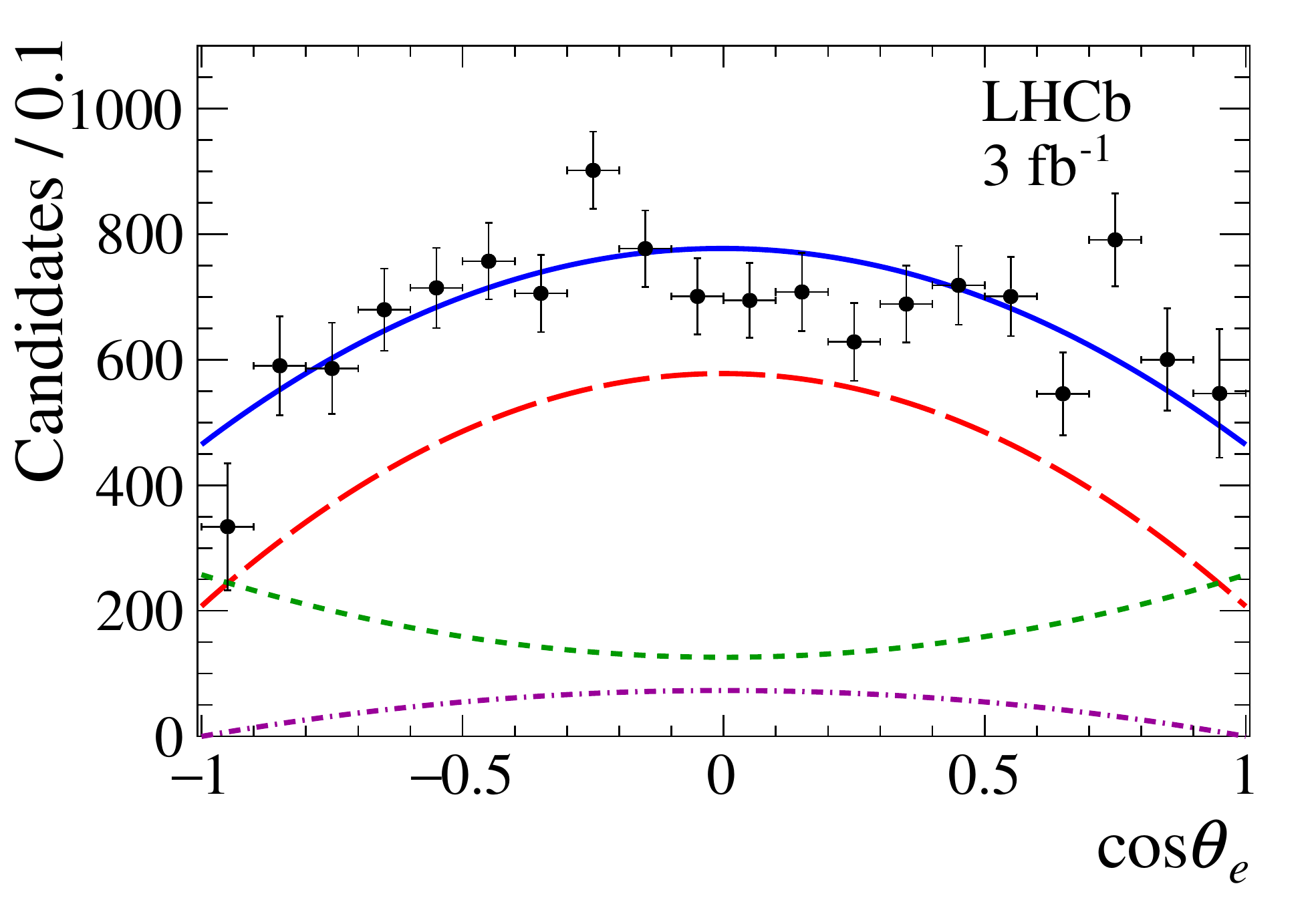} \\
 \includegraphics[width=0.4\linewidth]{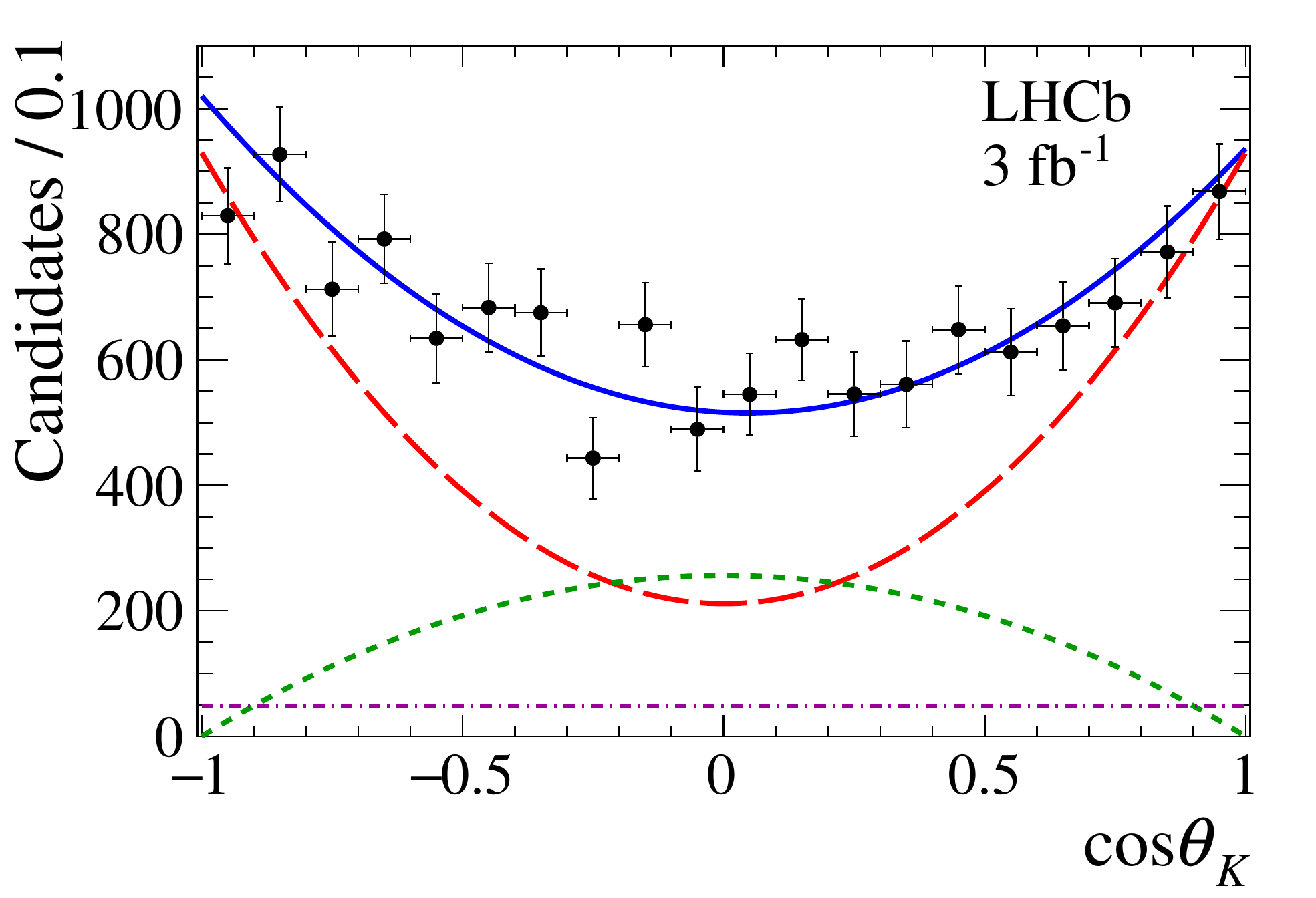}\includegraphics[width=0.4\linewidth]{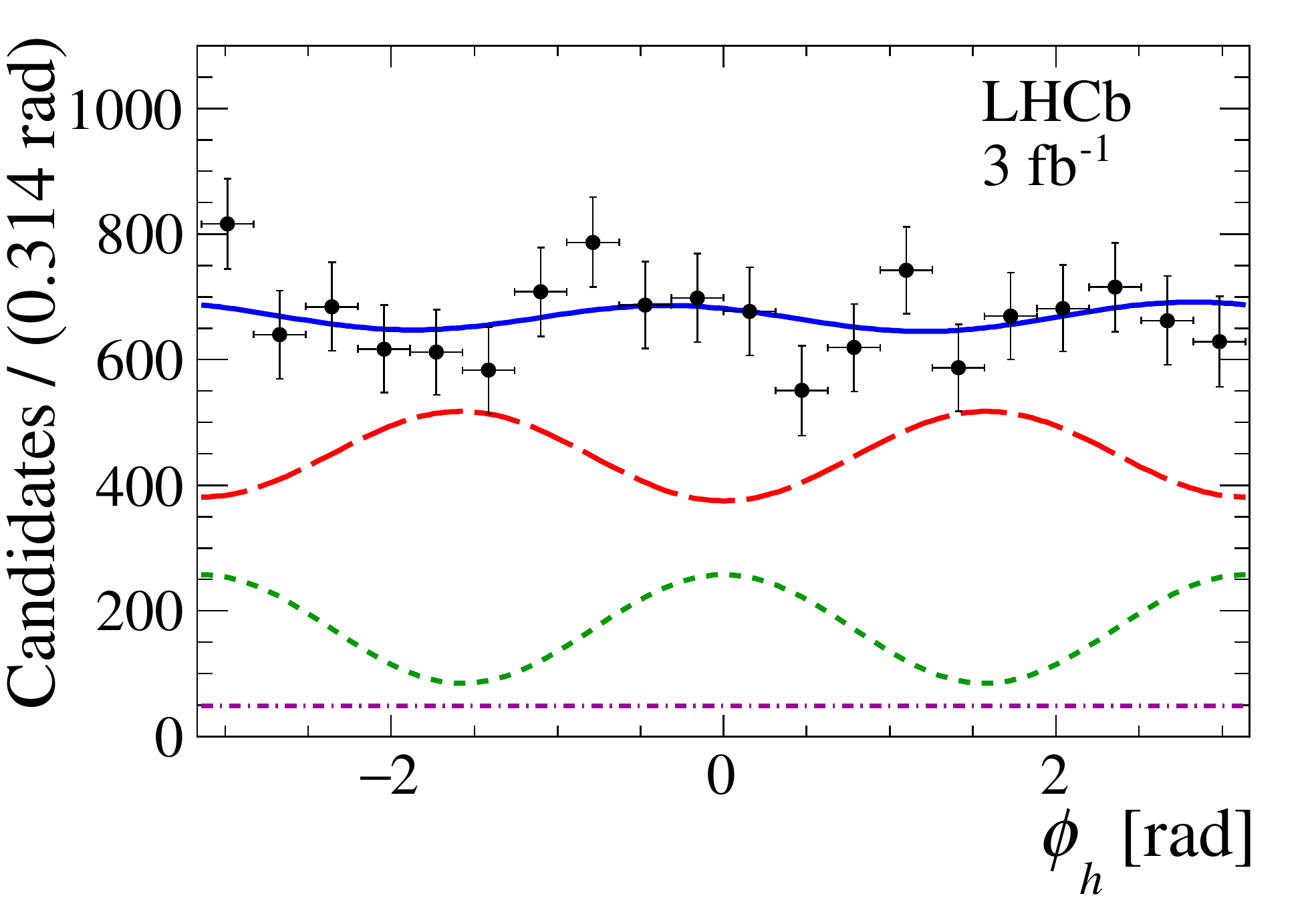} \\
     \vspace*{-0.5cm}
  \end{center}
    \caption{Decay time and helicity-angle distributions for (data points) \decay{\Bs}{\jpsi(}{\epem)\phiz} decays with the one-dimensional projections of the PDF extracted in the maximum-likelihood fit. The solid blue line shows the total signal contribution, which is composed of (long-dashed red) \CP-even, (short-dashed green) \CP-odd and (dash-dotted purple) S-wave contributions.}  
  \label{fig:TimeAngleFit} 
\end{figure}

The results of the fit to the data are shown in Table~\ref{tab:FitResults} while the projections of the fit results on the decay time and helicity-angle distributions are reported in Fig.~\ref{fig:TimeAngleFit}. The correlation matrix of statistical uncertainties is reported in Table~\ref{tab:CorrelationMatrix} of Appendix~\ref{sec:CorrelationMatrix}.
The results are consistent with previous measurements of these parameters~\cite{LHCb-PAPER-2019-013, Aad:2014cqa, Aad:2020jfw, Khachatryan:2015nza, Aaltonen:2012ie, Abazov:2011ry}, and the SM predictions for \phis~\cite{CKMfitter2005, Lenz:2006hd, Artuso:2015swg}. They show no evidence of \CP violation in the interference between \Bs meson mixing and decay, nor for direct \CP violation in \decay{\Bs}{\jpsi(}{\epem)\phiz} decays, as the parameter $|\lambda|$ is consistent with unity within uncertainties.

\begin{table}[t]
  \caption{Results of the maximum-likelihood fit, described in Sec.~\ref{sec:Results},  to the \decay{\Bs}{\jpsi(}{\epem)\phiz} decays including all acceptance and resolution effects. The first uncertainty is statistical and the second is systematic.}
\begin{center}\vspace*{-0.5cm}\begin{tabular}{ll}
    \toprule
Parameter & Fit result and uncertainty \\
\midrule
        $\Gs$ [\invps] &$0.608\pm0.018\pm0.012$ \\
        $\DGs$ [\invps] & $0.115\pm0.045\pm0.011$               \\
$|A_{\hspace{-1pt}\perp}|^{2}$ & $0.234\pm0.034\pm0.008$        \\
        $|A_0|^2$ & $0.530\pm0.029\pm0.013$ \\
$\delta_\parallel$ [\rad] & $3.11^{\,+\,0.08}_{\,-\,0.07}\pm0.06$\\
    $\delta_\perp$ [\rad] & $2.41^{\,+\,0.43}_{\,-\,0.42}\pm0.10$\\
            $\phis$ [\rad] & $0.00\pm0.28\pm0.07$               \\
       $|\lambda|$ & $0.877^{\,+\,0.112}_{\,-\,0.116}\pm0.031$  \\
    $F_\mathrm{S}$ & $0.062^{\,+\,0.042}_{\,-\,0.051}\pm0.022$    \\
    $\delta_\mathrm{S}$ [\rad] & $0.01^{\,+\,0.25}_{\,-\,0.27}\pm0.04$  \\
    \bottomrule
  \end{tabular}\end{center}
\label{tab:FitResults}
\end{table}

\section{Systematic uncertainties}
\label{sec:SystUnc}

Systematic uncertainties for each of the measured parameters are reported in Table~\ref{tab:SystError}. They are evaluated by observing the change in the physics parameters after repeating the likelihood fit with a modified model assumption, or through pseudoexperiments, in case of uncertainties originating from the limited size of calibration samples. 

The decay-time and angular efficiencies obtained independently in the three bremsstrahlung categories are compatible within statistical uncertainties. While the effective decay-time resolution differs for the three categories, it was verified with simulations that the result of a weighted average of three independent maximum-likelihood fits is consistent with the default one.

Repeating the mass fit in bins of the decay time and helicity angles shows that the mass resolution depends on $\cos\theta_e$ and $\cos\theta_K$. 
As the \sPlot technique assumes that the discriminating variable is independent of the observables of interest, the effect of this correlation is quantified.
The data sample is divided in intervals of $\cos\theta_e$ and $\cos\theta_K$ and new weights are computed with fits to $m(\epem\Kp\Km)$.
The four-dimensional likelihood fit is evaluated with modified weights.
The variation of each physics parameter is assigned as a systematic uncertainty.
For the decay time and azimuthal $\phi_h$ angle the effect is negligible.

The mass model is tested in two ways. First new sets of weights are computed using alternate PDF models. One set with the signal component of the $m(\epem\Kp\Km)$ distribution described by a sum of two Ipatia functions~\cite{Santos:2013gra}. Second set with the combinatorial background described by a second order Chebyshev polynomial. Third set with the combinatorial background described by an exponential function with slope fixed to an average value from samples with one and both electrons corrected for bremsstrahlung.
For the second test a set of pseudoexperiments is used by fluctuating the default mass model parameters within their uncertainties (accounting for correlations), providing a new set of weights.
The width of the obtained physics parameters distributions from the pseudoexperiments or the difference between the default and alternate PDF results is assigned as systematic uncertainty, whichever is larger.

\begin{table}[t]
\caption{Statistical and systematic uncertainties. A dash corresponds to systematic uncertainties that are negligible. Systematic uncertainties from different sources are added in quadrature.}
\begin{center}\vspace*{-0.5cm}
\scalebox{0.9}{
\begin{tabular}{lcccccccccc}
\toprule
Source             &\Gs & \DGs& $A_{\hspace{-1pt}\perp}^{2}$&$A_0^2$&$\delta_\parallel$&$\delta_\perp$&$\phis$&$|\lambda|$&$F_S$ &$\delta_S$\\
                   &[\ensuremath{\ps^{-1}}]&[\ensuremath{\ps^{-1}}]&                      &       &[\rad]            &[\rad]        & [\rad]   &           &      & [\rad]   \\
\midrule
Stat. uncertainty           &0.018  & 0.045 & 0.034 & 0.029 &$^{\,+\,0.08}_{\,-\,0.07}$&$^{\,+\,0.43}_{\,-\,0.42}$&0.28&$^{\,+\,0.112}_{\,-\,0.116}$&$^{\,+\,0.042}_{\,-\,0.051}$&$^{\,+\,0.25}_{\,-\,0.27}$  \\
\midrule
Mass factorisation          & $0.003$ & $0.003$ & $0.005$ & $0.007$ & $0.01$ & $0.03$ & $0.02$ & $0.011$ & $0.017$ & $0.01$ \\ 
Mass model         		    & $0.011$ & $0.005$ & $0.004$ & $0.005$ & $0.02$ & $0.14$ & $0.05$ & $0.011$ & $0.007$ & $0.04$ \\ 
Ang. acceptance  		    & $-$    & $-$    & $0.002$ & $0.001$ & $-$   & $0.02$ & $0.01$ & $0.005$ & $0.003$ & $0.02$ \\ 
Time resolution       		    & $0.002$ & $0.008$ & $0.004$ & $0.002$ & $0.06$ & $0.02$ & $0.03$ & $0.003$ & $0.002$ & $0.01$ \\ 
Time acceptance		    & $0.003$ & $0.003$ & $0.001$ & $0.001$ & $-$   & $-$   & $-$   & $0.001$ & $-$    & $-$   \\ 
MC (time acc.)              & $0.001$ & $0.001$ & $0.001$ & $-$    & $-$   & $-$   & $-$   & $-$    & $-$    & $-$   \\
MC (ang. acc.)               & $-$    & $-$    & $0.001$ & $0.001$ & $0.01$ & $0.01$ & $0.02$ & $0.017$ & $0.003$ & $-$   \\
$\Lb$ background   		    & $0.001$ & $0.001$ & $0.001$ & $0.001$ & $0.01$ & $-$   & $0.01$ & $0.005$ & $0.01\ \, $ & $-$   \\ 
Ang. resolution 		    & $-$    & $0.002$ & $0.002$ & $0.003$ & $-$   & $0.01$ & $-$   & $-$    & $0.005$ & $-$   \\ 
$B_c^+$ background	        & $0.003$ & $-$    & $-$    & $-$    & $-$   & $-$   & $-$   & $-$    & $-$    & $-$   \\ 
Fit bias           		    & $-$    & $-$    & $-$    & $0.009$ & $-$   & $-$   & $-$   & $0.020$ & $-$    & $-$   \\ 
\midrule
Syst. uncertainty 		    & 0.012 & 0.011 & 0.008 & 0.013 & 0.07 & 0.15 & 0.07 & 0.031 & 0.022 & 0.05 \\
\midrule
Total uncertainty           & 0.022 & 0.046 & 0.035 & 0.032 & 0.10 &$^{\,+\,0.46}_{\,-\,0.45}$& 0.29 &$^{\,+\,0.117}_{\,-\,0.121}$&$^{\,+\,0.047}_{\,-\,0.056}$&$^{\,+\,0.26}_{\,-\,0.28}$  \\
\bottomrule
\end{tabular}}\end{center}
\label{tab:SystError}
\end{table} 

The statistical uncertainty on the angular efficiency is propagated by repeating the fit using new sets of the ten integrals, $I_k$, systematically varied according to their covariance matrix.
The width of the obtained distributions for each physics parameter is taken as the systematic uncertainty.
The angular resolution is neglected in the maximum-likelihood fit.
The effect of this assumption is studied using pseudoexperiments, where the helicity angles are smeared according to the experimental resolution.
There is a small effect on the polarisation amplitudes, strong phase and decay width difference while all other parameters are unaffected.

A systematic contribution is evaluated to take into account the effect of the finite decay-time resolution by comparing pseudoexperiments with fixed and constrained decay-time resolution parameters. A sample of pseudoexperiments with the four-dimensional $\Bs\to\jpsi(\epem)\phi$ PDF including time and angular efficiencies is used. The procedure is evaluated for two scenarios: the former with decay-time resolution parameters fixed to generated values, and the latter with parameters constrained to twice the difference between values obtained from signal simulation with \decay{\jpsi}{\epem} and \decay{\jpsi}{\mumu} decays. The quadratic difference between the uncertainties of pseudoexperiments with fixed and constrained parameters is assigned as a systematic uncertainty.
In addition tests with decay-time resolution parameters fixed in the fit to the data sample are performed.
The parameters are fixed to values obtained from the time angle fit at $\phis$ value fixed to 0 or $\pi/2$, or to values from a sample of \decay{\jpsi}{\mumu} candidates produced at the PV corrected for the difference between \epem and \mumu simulation samples.
The test results are compatible within statistical uncertainties to the default fit results.

The decay-time efficiency introduces a systematic uncertainty from three different sources. First, the contribution due to the statistical uncertainty on the determination of the decay-time efficiency from the control channel is obtained by evaluating the fit multiple times after randomly varying the parameters of the time efficiency within their statistical uncertainties. The statistical uncertainty is dominated by the size of the \decay{\Bd}{\jpsi\Kstar(892)^0} control sample. Second, a sum of two Ipatia functions is used as an alternative mass model for the $m(\epem\Kp\pim)$ distribution and a new decay-time efficiency function is produced. Finally, the efficiency function is computed with the \Bd lifetime modified by $\pm1\sigma$. In all cases the difference in the fit results arising from the use of the new efficiency function is taken as a systematic uncertainty.

The sensitivity to the BDT selection is studied by adjusting the working point around the optimal position for the signal channel where the difference of the number of signal candidates is within $10\%$ between the default and varied BDT criteria.
The effect of applying the modified BDT requirement in the likelihood fit is studied using pseudoexperiments.
The mass model parameters for each BDT requirement are varied within their uncertainties (accounting for correlations) and the weights are re-evaluated based on the alternative model.
The fit is repeated using a new set of weights and a new efficiency function.
The observed variations in the physics parameters are compatible with statistical fluctuations.
This is verified by pseudoexperiments with $10\%$ of candidates removed at random. 

A systematic uncertainty is assigned to account for the differences in the final-state kinematics between data and simulated samples.
The simulated signal events are weighted using a multidimensional BDT-based algorithm~\cite{Rogozhnikov:2016bdp} in six dimensions corresponding to kinematic variables with largest observed discrepancies between data and simulations.
The procedure is repeated for the control sample $\Bd\to\jpsi(\epem)\Kstar(892)^0$.
The reweighted simulation samples of both channels are used to obtain new angular and decay-time acceptances.
The difference with the default fit result is assigned as a systematic uncertainty.

The fraction of \decay{\Lb}{\jpsi p\Km} candidates contributing to the signal sample is estimated to be $1\%$ using simulation.
The impact of neglecting this contribution is evaluated for the data sample by fitting the $m(\epem\Kp\Km)$ distribution with an additional component to account for, namely the sum of two CB functions, the shape of which is fixed to a fit to simulated \decay{\Lb}{\jpsi p\Km} candidates.
In addition, the decay-time efficiency is redetermined including a component for background from \decay{\Lb}{\jpsi p\pim} decays. This component is modelled by the sum of two CB functions, the shape of which is fixed to a fit to simulated \decay{\Lb}{\jpsi p\pim} candidates. The fraction of the \decay{\Lb}{\jpsi p\pim} decays is estimated from the simulation to be at most $0.06\%$~\cite{LHCb-PAPER-2015-032}. 
The differences of physics parameters obtained from the fit with modified weights and efficiency function is assigned as a systematic uncertainty.

A small fraction of \decay{\Bs}{\jpsi\phiz} decays comes from the decays of \Bc mesons. The fraction is estimated as $0.8\%$ in Ref.~\cite{LHCb-PAPER-2013-044} and pseudoexperiments are used to assess the impact of ignoring such a contribution on the extraction of the physics parameters. Only \Gs is observed to be affected, with a bias on its central value corresponding to $20\%$ of the statistical uncertainty, which is assigned as a systematic uncertainty.

A possible bias in the fitting procedure is investigated through many pseudoexperiments of equivalent size to the data sample. 
For each pseudoexperiment the physics parameters are fluctuated in the underlying PDF and then compared to the obtained fit results. 
The resulting deviations are small and those that are not compatible with zero within three standard deviations are quoted as systematic uncertainties.

Inclusion of a result with a constraint on the $\dms$ into a global analysis leads to troublesome treatment of systematic effects introduced by choice of the constraint.
Therefore we provide a result with the mixing frequency fixed to the PDG value, \mbox{$\dms=17.757\invps$}~\cite{PDG2020}, as reported in Appendix~\ref{sec:FixedDms}. No significant difference is observed with respect to the default result.

The systematic uncertainties associated to the mass model and mass factorisation can be treated as uncorrelated between this result and that of Ref.~\cite{LHCb-PAPER-2014-059}.
More details on the systematic effects for the studied channel are given in Ref.~\cite{Batozskaya:2751174}. 


\section{Conclusion}
\label{sec:Concl}

Using a data set corresponding to an integrated luminosity of $3 \invfb$ collected by the \lhcb experiment in $pp$ collisions 
at centre-of-mass energies of $7$ and $8\tev$, a flavour-tagged decay-time-dependent angular analysis of $(1.27\pm0.05)\times10^4$ \decay{\Bs}{\jpsi(}{\epem)\phiz} decays is performed.
A number of physics parameters including the \CP-violating phase \phis, average decay width \Gs and decay width difference \DGs as well as the polarisation amplitudes and strong phases of the decay are determined. The effective decay-time resolution and effective tagging power are $45.6\pm0.1\fs$ and $(5.07\pm0.16)\%$, respectively.
The \CP parameters are measured to be

\begin{equation*}
\begin{aligned}
  \phis & = 0.00\pm0.28\pm0.07\rad,\\
  \DGs  & = 0.115\pm0.045\pm0.011\invps,\\
  \Gs   & = 0.608\pm0.018\pm0.012\invps
\end{aligned}
\end{equation*}

where the first uncertainty is statistical and the second is systematic.
The dominant sources of the systematic uncertainty are the imperfect mass and decay-time resolution models. This is the first measurement of the \CP content of the \decay{\Bs}{\jpsi(}{\epem)\phiz} decay and first time that \phis has been measured in the final state containing electrons.
These results constitute an important check for the results with muons in the final state because the systematic uncertainties of the measurements are independent, while the studied mechanism of the \CP violation is the same.
The results are consistent with previous measurements~\cite{LHCb-PAPER-2019-013, Aad:2014cqa, Aad:2020jfw, Khachatryan:2015nza, Aaltonen:2012ie, Abazov:2011ry}, the SM predictions~\cite{CKMfitter2005, Lenz:2006hd, Artuso:2015swg}, and show no evidence of \CP violation in the interference between \Bs meson mixing and decay. 
In addition, no evidence for direct \CP violation in \decay{\Bs}{\jpsi(}{\epem)\phiz} decays is observed.

\section*{Acknowledgements}
%
%
\noindent We express our gratitude to our colleagues in the CERN
accelerator departments for the excellent performance of the LHC. We
thank the technical and administrative staff at the LHCb
institutes.
We acknowledge support from CERN and from the national agencies:
CAPES, CNPq, FAPERJ and FINEP (Brazil); 
MOST and NSFC (China); 
CNRS/IN2P3 (France); 
BMBF, DFG and MPG (Germany); 
INFN (Italy); 
NWO (Netherlands); 
MNiSW and NCN (Poland); 
MEN/IFA (Romania); 
MSHE (Russia); 
MICINN (Spain); 
SNSF and SER (Switzerland); 
NASU (Ukraine); 
STFC (United Kingdom); 
DOE NP and NSF (USA).
We acknowledge the computing resources that are provided by CERN, IN2P3
(France), KIT and DESY (Germany), INFN (Italy), SURF (Netherlands),
PIC (Spain), GridPP (United Kingdom), RRCKI and Yandex
LLC (Russia), CSCS (Switzerland), IFIN-HH (Romania), CBPF (Brazil),
PL-GRID (Poland) and NERSC (USA).
We are indebted to the communities behind the multiple open-source
software packages on which we depend.
Individual groups or members have received support from
ARC and ARDC (Australia);
AvH Foundation (Germany);
EPLANET, Marie Sk\l{}odowska-Curie Actions and ERC (European Union);
A*MIDEX, ANR, Labex P2IO and OCEVU, and R\'{e}gion Auvergne-Rh\^{o}ne-Alpes (France);
Key Research Program of Frontier Sciences of CAS, CAS PIFI, CAS CCEPP, 
Fundamental Research Funds for the Central Universities, 
and Sci. \& Tech. Program of Guangzhou (China);
RFBR, RSF and Yandex LLC (Russia);
GVA, XuntaGal and GENCAT (Spain);
the Leverhulme Trust, the Royal Society
 and UKRI (United Kingdom).

\section*{Appendices}

\appendix

\section{Correlation matrix}
\label{sec:CorrelationMatrix}

The \CP observables are determined by an unbinned maximum-likelihood fit to the background-subtracted candidates with a probability density function (PDF) describing \decay{\Bs}{\jpsi(}{\epem)\phiz} signal decay. The correlation matrix of their statistical uncertainties is presented in Table~\ref{tab:CorrelationMatrix}. It is obtained using the bootstrap method.

\begin{table}[!htbp]
  \caption{Correlation matrix of statistical uncertainties.} 
\begin{center}\begin{tabular}{lrrrrrrrrrr}
\toprule
& $\Gs$ & $\DGs$ & $|A_{\hspace{-1pt}\perp}|^2$ & $|A_0|^2$ & $\delta_\parallel$ & $\delta_\perp$ & $\phis$ & $|\lambda|$ & $F_S$ & $\delta_S$ \\ \midrule
$\Gs$                          & $1.00$ & $-0.31$ & $0.41$        & $-0.38$       & $-0.01$  & $-0.03$ & $0.0$ & $-0.09$ & $-0.08$ & $-0.03$  \\
$\DGs$                         &      & $1.00$  & $-0.68$ & $0.63$  & $0.01$  & $-0.02$ & $0.01$  & $-0.01$ & $-0.04$ & $-0.02$ \\
$|A_{\hspace{-1pt}\perp}|^{2}$ &      &       & $1.00$        & $-0.66$ & $-0.06$ & $0.10$  & $-0.06$ & $-0.14$ & $-0.26$ & $0.03$  \\
$|A_0|^2$                      &      &       &             & $1.00$        & $0.08$  & $-0.17$ & $0.07$  & $0.24$  & $0.36$  & $-0.05$ \\
$\delta_\parallel$             &      &       &             &             & $1.00$  & $-0.03$ & $0.13$  & $-0.06$ & $0.14$  & $-0.20$ \\
$\delta_\perp$                 &      &       &             &             &       & $1.00$  & $0.08$  & $-0.11$  & $-0.28$ & $-0.05$ \\
$\phi_s$                       &      &       &             &             &       &       & $1.00$  & $0.15$  & $0.26$  & $-0.05$ \\
$|\lambda|$                    &      &       &             &             &       &       &       & $1.00$  & $0.52$  & $-0.03$  \\
$F_S$                          &      &       &             &             &       &       &       &       & $1.00$  & $-0.06$ \\
$\delta_S$                     &      &       &             &             &       &       &       &       &       & $1.00$  \\
\bottomrule
\end{tabular}\end{center}
\label{tab:CorrelationMatrix}
\end{table}

\section{Fit results with fixed \texorpdfstring{\dms}{dms}}
\label{sec:FixedDms}

The fit is repeated with a fixed value of the mixing frequency \mbox{$\dms=17.757\invps$}~\cite{PDG2020} instead of a Gaussian constraint.
The fit results are presented in Table~\ref{tab:FitResultsFixedDms} and corresponding correlation matrix in Table~\ref{tab:CorrelationMatrixFixedDms}.

\begin{table}[!htbp]
  \caption{Results of the maximum-likelihood fit described in Section~\ref{sec:Results} to the \decay{\Bs}{\jpsi(}{\epem)\phiz} decays including all acceptance and resolution effects and with the mixing frequency fixed to the PDG value, \mbox{$\dms=17.757\invps$} ~\cite{PDG2020}. The first uncertainty is statistical and the second is systematic, which is discussed in Section~\ref{sec:SystUnc}.}
\begin{center} \begin{tabular}{ll}
    \toprule
Parameter & Fit result and uncertainty \\
\midrule
                $\Gs$ [\invps] & $0.608\pm0.018\pm0.012$    \\
               $\DGs$ [\invps] & $0.115\pm0.043\pm0.011$  \\
$|A_{\hspace{-1pt}\perp}|^{2}$ & $0.234\pm0.033\pm0.008$      \\
                     $|A_0|^2$ &$0.53^{\,+\,0.026}_{\,-\,0.027}\pm0.013$ \\
     $\delta_\parallel$ [\rad] & $3.11^{\,+\,0.07}_{\,-\,0.08}\pm0.07$\\
         $\delta_\perp$ [\rad] & $2.41^{\,+\,0.45}_{\,-\,0.46}\pm0.15$\\
                $\phis$ [\rad] & $0.00\pm0.30\pm0.07$           \\
                   $|\lambda|$&$0.877^{\,+\,0.104}_{\,-\,0.126}\pm0.031$\\
                $F_\mathrm{S}$&$0.062^{\,+\,0.045}_{\,-\,0.052}\pm0.022$\\
    $\delta_\mathrm{S}$ [\rad] & $0.01\pm0.29\pm0.05$               \\
    \bottomrule
  \end{tabular}\end{center}
\label{tab:FitResultsFixedDms}
\end{table}

\begin{table}[!htbp]
  \caption{Correlation matrix of statistical uncertainties for a fit with fixed \dms.} 
\begin{center}\begin{tabular}{lrrrrrrrrrr}
\toprule
& $\Gs$ & $\DGs$ & $|A_{\hspace{-1pt}\perp}|^2$ & $|A_0|^2$ & $\delta_\parallel$ & $\delta_\perp$ & $\phis$ & $|\lambda|$ & $F_S$ & $\delta_S$ \\ \midrule
$\Gs$                          & $1.00$ & $-0.25$ & $0.37$        & $-0.32$       & $-0.01$  & $-0.04$ & $0.01$ & $-0.04$ & $-0.05$ & $-0.03$  \\
$\DGs$                         &      & $1.00$  & $-0.65$ & $0.60$  & $0.04$  & $-0.05$ & $0.06$  & $-0.01$ & $-0.09$ & $0.05$ \\
$|A_{\hspace{-1pt}\perp}|^{2}$ &      &       & $1.00$        & $-0.61$ & $-0.13$ & $0.09$  & $-0.09$ & $-0.17$ & $-0.25$ & $0.01$  \\
$|A_0|^2$                      &      &       &             & $1.00$        & $0.14$  & $-0.16$ & $0.07$  & $0.17$  & $0.31$  & $0.0$ \\
$\delta_\parallel$             &      &       &             &             & $1.00$  & $-0.05$ & $0.10$  & $-0.01$ & $0.17$  & $-0.22$ \\
$\delta_\perp$                 &      &       &             &             &       & $1.00$  & $0.20$  & $-0.07$  & $-0.26$ & $-0.10$ \\
$\phi_s$                       &      &       &             &             &       &       & $1.00$  & $0.20$  & $0.20$  & $-0.07$ \\
$|\lambda|$                    &      &       &             &             &       &       &       & $1.00$  & $0.51$  & $-0.03$  \\
$F_S$                          &      &       &             &             &       &       &       &       & $1.00$  & $-0.05$ \\
$\delta_S$                     &      &       &             &             &       &       &       &       &       & $1.00$  \\
\bottomrule
\end{tabular}\end{center}
\label{tab:CorrelationMatrixFixedDms}
\end{table}


\addcontentsline{toc}{section}{References}
\bibliographystyle{LHCb}
\bibliography{main,standard,LHCb-PAPER,LHCb-CONF,LHCb-DP,LHCb-TDR}

\newpage
\centerline
{\large\bf LHCb collaboration}
\begin
{flushleft}
\small
R.~Aaij$^{31}$,
C.~Abell{\'a}n~Beteta$^{49}$,
T.~Ackernley$^{59}$,
B.~Adeva$^{45}$,
M.~Adinolfi$^{53}$,
H.~Afsharnia$^{9}$,
C.A.~Aidala$^{84}$,
S.~Aiola$^{24}$,
Z.~Ajaltouni$^{9}$,
S.~Akar$^{64}$
% ...

{\footnotesize \it

% ...
$^{9}$Universit{\'e} Clermont Auvergne, CNRS/IN2P3, LPC, Clermont-Ferrand, France\\
% ...
$^{24}$INFN Sezione di Milano, Milano, Italy\\
% ...
$^{31}$Nikhef National Institute for Subatomic Physics, Amsterdam, Netherlands\\
% ...
$^{45}$Instituto Galego de F{\'\i}sica de Altas Enerx{\'\i}as (IGFAE), Universidade de Santiago de Compostela, Santiago de Compostela, Spain\\
% ...
$^{49}$Physik-Institut, Universit{\"a}t Z{\"u}rich, Z{\"u}rich, Switzerland\\
% ...
$^{53}$H.H. Wills Physics Laboratory, University of Bristol, Bristol, United Kingdom\\
% ...
$^{59}$Oliver Lodge Laboratory, University of Liverpool, Liverpool, United Kingdom\\
% ...
$^{64}$University of Cincinnati, Cincinnati, OH, United States\\
% ...
$^{67}$Syracuse University, Syracuse, NY, United States\\
% ...
$^{84}$University of Michigan, Ann Arbor, United States, associated to $^{67}$\\
\bigskip
% ...
\medskip
}
\end{flushleft}




\end{document}